\documentclass[12pt]{article}
\usepackage{graphicx}
\usepackage{amsmath}
\usepackage{url}
\usepackage{courier}
\usepackage{xcolor}
\newcommand{\be}{\begin{equation}}
\newcommand{\ee}{\end{equation}}
\begin{document}
\begin{titlepage}
\title{{\bf Dalitz-plot analysis of $B^{\pm} \to K^{\pm}K^+K^-$ decays}}
\author{
{L. Le\'sniak\footnote{email: leonard.lesniak@ifj.edu.pl} ~and P. \.Zenczykowski}\\
{\it Division of Theoretical Physics}\\
{\it The Henryk Niewodnicza\'nski
Institute of Nuclear Physics,}\\
{\it Polish Academy of Sciences}\\
{\it PL 31-342 Krak\'ow, Poland}
}
\maketitle
\begin{abstract}
We study
$B^{\pm} \to K^{\pm}K^+K^-$ decays  
using the QCD factorization model
with final state interactions between $K^+$ and $K^-$ mesons taken into account.
The parameters of the model are fitted to the data of the {$ BABAR$} and LHCb collaborations.
We describe the $K \bar{K}$ effective mass distributions and examine the $CP$-violating asymmetry effects in the full range of the Dalitz plot.
 
\end{abstract}
PACS numbers: 13.25.Hw, 13.75.Lb
\vfill

\end{titlepage}

\section{Introduction} \label{Introduction}

Theoretical understanding of $CP$ violation effects in $B$ weak decays
requires proper description of both weak and strong phases of the decay amplitudes.
The latter are in principle determined by the short- and long-distance QCD interactions. 
While the two-body
$B$ decays provide the simplest setting for the study of the interplay of weak
and strong interactions, their three-body decays call for the analysis
of more complicated  strong interaction effects that may lead
to significant variability of $CP$-asymmetry in the Dalitz plots.
With the relevant effects
being induced by various long-distance contributions (such as final state interactions (FSI)),
their description is difficult and requires the use of QCD-inspired models.

Of particular interest are here the three-body $B$ decays into light mesons
such as $B \to \pi\pi\pi,~\pi\pi K,~\pi KK,~KKK$. The long-distance strong interactions
include both the  resonant and nonresonant contributions
(see \cite{Cheng2013},\cite{Cheng2016b}) that may be mutually interfering and could be described
as proceeding through a quasi-two-body stage $B \to M_1 M_2$ with meson $M_2$ ultimately decaying into two
final-state mesons $m$ and $m'$. 
The description of the relevant induced invariant-mass-dependent effects calls for the careful
consideration of both the resonances and their interferences in the low-mass regions of the Dalitz plot
such as $\rho(770)$, $\omega(782)$, $f_0(980)$, $\phi(1020)$, $f_2(1270)$, $\omega(1420)$, $\rho(1450)$, $f^{'}_2(1525)$,
\textit{etc.} (\cite{Zou2020}-\cite{Zou2021})
 as well as various hadronic loop or penguin contributions.
The latter may involve the  $\pi\pi \to KK$ rescattering as discussed in 
\cite{Furman2005}-\cite{Alvarez}
and 
the effects of charm $D\bar{D}$ threshold opening up
at higher invariant masses \cite{Bediaga2018},\cite{Mannel2020}.

In our work in the past (see \cite{Furman2005}, \cite{FKLZ}, \cite{LZ14}) we concentrated on an analysis of the $B^{\pm} \to K^{\pm}K^+K^-$ decays \cite{Belle}-\cite{LHCb13},  
using the quasi-two-body QCD factorization model \cite{rhonumber} 
with some FSI added.
The $K^+K^-$ rescattering effects were restricted to fairly low values of the $K{\bar K}$ invariant
mass.  The availability of the experimental data in the region of
the large effective mass values of the interacting $K^+K^-$ pair,
and a large variability of $CP$ asymmetry in the Dalitz plot 
as observed by {$ BABAR$} and LHCb \cite{BABAR12},\cite{LHCb14},\cite{LHCb16}, are important as they put stringent constraints on the theoretical models.
It is therefore challenging to extend our previous work and
describe the $B^{\pm} \to K^{\pm}K^+K^-$ decays in the full range of the Dalitz plot.
Such an extension of our model constitutes the subject of the present paper.

We present our extended model in Sec. 2 where the relevant amplitudes are defined.
The Dalitz plot distributions are discussed in Sec. 3. Various parameters and
the fitting procedure are described in Sec. 4. Our results are given in Sec. 5.
Summary and final comments can be found in Sec. 6.

\section{Amplitudes of the $B^{\pm} \to K^{\pm}K^+K^-$ decays } \label{Amplitudes}

In this chapter we widen our studies of the $B^{\pm} \to K^{\pm}K^+K^-$ decays presented in Refs.~\cite{Furman2005}, ~\cite{FKLZ} and ~\cite{LZ14}.
  We employ the quasi-two-body QCD factorisation
model \cite{rhonumber} of the decay
                             amplitudes extending it to large effective
$K\bar{K}$ masses. In order to describe
                             the full range of these masses, up to about
4.78 GeV, the relevant decay
                             amplitudes have to be modified. In this section we give
general theoretical expressions for the
                             decay amplitudes. Our modifications will be
introduced in the subsequent
sections.\\

The amplitude $A^-(s_{12},s_{23})$ for the $B^{-} \to K^{-}(p_1)K^+(p_2)K^-(p_3)$ decay is a function of two $K\bar{K}$ invariant masses squared: $s_{12}=(p_1+p_2)^2$ and $s_{23}=(p_2+p_3)^2$, where
$p_1$, $p_2$ and $p_3$ are the kaon four-momenta.
This decay amplitude has to be symmetrized as in the final state two kaons of negative charge are emitted.  
The symmetrized amplitude $A^-_{sym}$ is 
\be
\label{A-sym}A^-_{sym}=\frac{1}{\sqrt{2}} [A^-(s_{12},s_{23})+A^-(s_{23},s_{12})].
\ee
The amplitude $A^-(s_{12},s_{23})$ is a sum of six components $A^-_i$:

\be
\label{A-}
A^-(s_{12},s_{23})=\sum_{i=1}^{6} A^-_i,
\ee
where
\begin{equation}
\label{A1}
A^-_1=\frac{G_F}{\sqrt{2}}\,\nu\,\frac{2B_0}{m_b-m_s}(M^2_B-m^2_K)F_0^{B^-K^-}(s_{23}) \,\Gamma^{s*}_2(s_{23}),
\end{equation}

\begin{equation}
\label{A2}
A^-_2=-\frac{G_F}{\sqrt{2}}\, y\,\sqrt{\frac{1}{2}}f_K (M^2_B-s_{23})
F_0^{B^-\to(K^+K^-)_S}(m^2_K)[\,\chi\,\Gamma^{n*}_2(s_{23})+G_1(s_{23})],
\end{equation}

\begin{equation}
\label{A3}
A^-_3=\frac{G_F}{\sqrt{2}}(s_{12}-s_{13})
F_1^{B^-K^-}(s_{23})\left[w_uF_u^{K^+K^-}(s_{23})+w_sF_s^{K^+K^-}(s_{23})\right],
\end{equation}

\be
\label{A4}
A^-_4=-\frac{G_F}{\sqrt{2}}\, y\,\frac{f_K}{f_{\rho}}\,(s_{12}-s_{13})A_0^{B^-\rho^0}(m^2_K)\,F_u^{K^+K^-}(s_{23}),
\end{equation}

\be
\label{A5}
A^-_5=-\frac{G_F}{\sqrt{2}}\, y\,f_K\, D(s_{12},s_{23})\,\langle f_2|u\bar{u}\rangle \,G_{f_2K^+K^-}(s_{23})\,  F^{B^-f_2}(m_K^2),
\ee
and
\be
\label{A6}
A^-_6=-\frac{G_F}{\sqrt{2}}\, y\,f_K\, D(s_{12},s_{23})\,\langle f_2^{'}|u\bar{u}\rangle \,
G_{{f_2} {^{'}} K^+K^-}(s_{23})\, 
F^{B^{-}{f_2}'}(m_K^2).
\ee

The amplitudes $A^-_1$ and $A^-_2$ depend on the $K^+K^-$ interactions in the $S$-wave, the $A^-_3$ and $A^-_4$ terms constitute the $P$-wave components, while the amplitudes $A^-_5$ and $A^-_6$ involve the $D$-wave resonances $f_2(1270)$ and $f_2^{'}(1525)$.
All the amplitudes are proportional to the Fermi coupling constant $G_F$.
They also depend on QCD factorization coefficients $a_j^c$ and $a_j^u$ (j=1,...,10), and on the products $\Lambda_u=V_{ub}V_{us}^*$, 
$\Lambda_c=V_{cb}V_{cs}^*$, where $V_{kl}$ are the CKM quark-mixing matrix elements.
These coefficients as well as $\Lambda_u$ and $\Lambda_c$ enter in the factors
$\nu$, $y$, $w_u$ and $w_s$ present in Eqs.~(\ref{A1})-(\ref{A6}). They are defined as follows:

\begin{equation}
 \label{v}
\nu=\Lambda_u(-a_{6\nu}^u+\dfrac{1}{2}a_{8\nu}^u)+\Lambda_c(-a_{6\nu}^c+
\dfrac{1}{2}a_{8\nu}^c),
\end{equation}

\begin{equation}
\label{y}
\begin{split}
 y=\Lambda_u\Big[a_{1y}+a_{4y}^u+a_{10y}^u-(a_{6y}^u
+a_{8y}^u)r_{\chi}^K\Big]+
\\
\Lambda_c\Big[a_{4y}^c+
a_{10y}^c-(a_{6y}^c+a_{8y}^c)r_{\chi}^K\Big],
\end{split}
\end{equation}
where
\begin{equation}
\label{rchi}
 r_{\chi}^K=\dfrac{2m_K^2}{(m_b+m_u)(m_u+m_s)},
\end{equation}

\begin{equation}
\label{wu}
 w_u=\Lambda_u(a_{2w}+a_{3w}+a_{5w}+a_{7w}+
a_{9w})+\Lambda_c(a_{3w}+a_{5w}+a_{7w}+a_{9w}),
\end{equation}
and
\begin{equation}
\label{ws}
\begin{split}
 w_s=\Lambda_u\Big[a_{3w}+a_{4w}^u+a_{5w}-\dfrac{1}{2}(a_{7w}
+a_{9w}+a_{10w}^u)\Big]+
\\
\Lambda_c\Big[a_{3w}+a_{4w}^c+
a_{5w}-\dfrac{1}{2}(a_{7w}+a_{9w}+a_{10w}^c)\Big].
\end{split}
\end{equation}

The masses of the $B^{\pm}$ and the $K^{\pm}$ mesons and those of the $b$-quark,
$u$-quark, $d$-quark and the strange quark are denoted as $M_B$, $m_K$, $m_b$, $m_u$, $m_d$, and $m_s$, respectively.
The values of the coefficients  $a_j$, $a_j^c$ and $a_j^u$, calculated at next-to-leading order, are given in Table 1 of Ref.~\cite{FKLZ}.

The constant $B_0$ in the amplitude $A_1^-$ (Eq.~\ref{A1}) is defined as $B_0=m_{\pi}^2/(m_u+m_d)$, where $m_{\pi}$ is the mass of the charged pion.
The function $F_0^{B^-K^-}(s_{23})$ is the $B^- \to K^-$ scalar transition form factor.
We follow Ref.~\cite{Ball} and parameterize it as  $F_0^{B^-K^-}(s)=r_0/(1-s/s_0)$, where $r_0=$ 0.33 and $s_0=$ 37.46 GeV$^2$. 

The function $\Gamma^{s}_2(s_{23})$ is the kaon strange scalar form factor.
In its calculation  
the knowledge of the strong $K \bar{K}$ interactions as well as the interactions in the meson channels coupled to $K^+ K^-$ (like the
$\pi \pi$ and four pion systems) is needed.
 A unitary model of the $S$-wave $\pi \pi$, $K \bar{K}$ and the 4$\pi$ amplitudes has been developed in Refs.~\cite{KLL1} and~\cite{KLL2}.
Recently, it has been updated using some new experimental data (see Appendix A in
Ref.~\cite{DKKK}). 
 The coupling of the four pion channel to the $K
\bar K$ channel becomes important when the corresponding $K^+K^-$ effective mass exceeds a value of about 1300 MeV.  
It is known that two scalar resonances $f_0(1370)$ and $f_0(1500)$ have large decay branching fractions into four pions~\cite{PDG22}.
The four pions can form clusters of two-pion systems like $\rho\rho$ or $\sigma\sigma$
and their interactions can lead to an enhancement of the $K\bar K$ mass distribution near 1.5 GeV.

Applying the above three-channel model and the Muskhelishvili-Omnes dispersion relations Bachir Moussallam~\cite{BM} has calculated the kaon strange scalar form factor $\Gamma^{s}_2(s_{23})$
and the non-strange scalar form factor $\Gamma^{n}_2(s_{23})$ which appears in the amplitude $A_2^-$ (Eq.~(\ref{A2})). 
At this stage some additional assumptions about an asymptotic behaviour of the phase shifts for the centre of mass energy above 1.8 GeV have been necessary since in that energy range no experimental data on the meson-meson interactions are available. 
Labelling by $i=1$ the pion-pion channel, by $i=2$ the $K\bar K$ channel and by $i=3$ the effective $2\pi2\pi$ channel the asymptotic limits of the phase shifts $\delta_{ii}$ have been taken as $2\pi$, $\pi$ and 0, respectively for $i$ equal to 1, 2 and 3.
\footnote{In Appendix A of Ref.~\cite{DKKK} these asymptotic limits have been taken as
$3\pi$, $\pi$ and $-\pi$ what changed a behaviour of the kaon form factors.
}
\begin{figure}
\begin{center}
 \includegraphics[width=8cm]{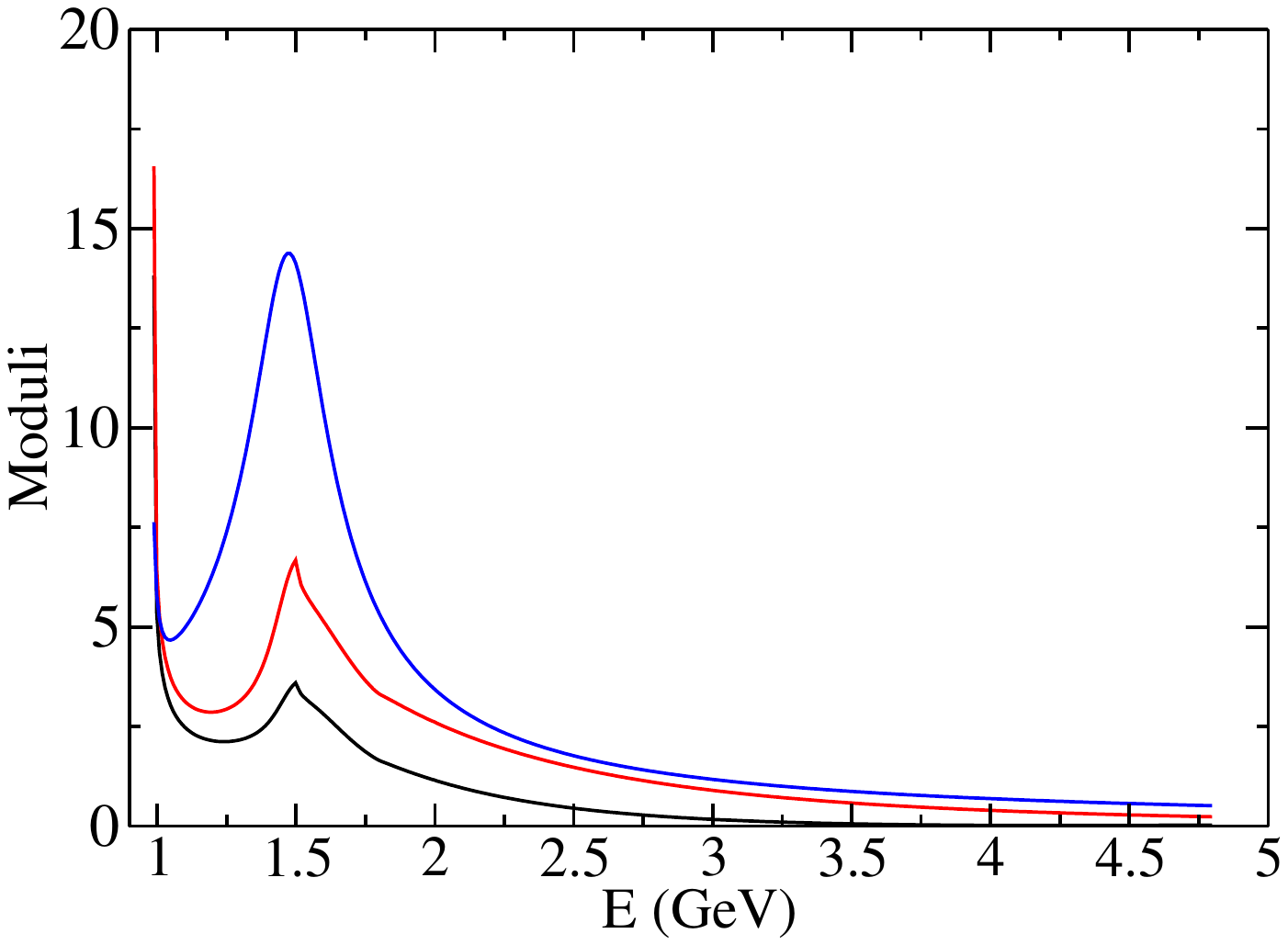}
\vspace*{6pt}
\caption{\label{fig1} Moduli of the kaon scalar strange form factor $\Gamma^{s}_2(s)$ (black line), 
the kaon scalar non-strange form factor $\Gamma^{n}_2(s)$ (red line), and the transition function $G_1(s)$ divided by the constant $\chi$ (blue line) plotted as functions of 
$E=\sqrt s$.}
\end{center}
\end{figure}
 
The amplitude $A^-_2$ involves
 the charged kaon decay constant $f_K$ and
the transition form factor $F_0^{B^-\to(K^+K^-)_S}(m^2_K)$ from the $B^-$ meson to the $K^+K^-$ pair in the $S$-wave.
As in Ref.~\cite{ElBennich} we take its value as 0.13.
One also has the kaon non-strange scalar form factor
$\Gamma^{n}_2(s_{23})$ multiplied by the constant $\chi$ related to the decay strength of the intermediate $S$-wave state into $K^+K^-$ pair.   
In comparison with a set of the $S$-wave amplitudes used in Ref.~\cite{LZ14} a new component of the $A^-_2$ amplitude proportional to the function $G_1(s_{23})$ appears in Eq.~(\ref{A2}).
In this component a transition from the $u \bar u$ pair of quarks into the $K^+K^-$ spin zero isospin one state is included. 
Its strength is given by a complex coupling constant $r_2$ which will be fitted to data.

The transition function $G_1(s_{23})$ has been defined in Ref.~\cite{DKKK} and used in the description of the $D^0 \to K^0_S K^+K^-$ decays where the $K^+K^-$ pair also appears in the final state. 
A coupling of the $S$-wave isospin one $\pi \eta$ state to the $K^+K^-$ state enters in a derivation of the transition function $G_1$.
The $\pi \eta$ and the $K^+K^-$ states have been introduced in a unitary way. The 
$G_1$ function has two poles related to the scalar-isovector resonances $a_0(980)$ and $a_0(1450)$.   
In the present application the ratio of the $\pi \eta$ coupling to the $K^+K^-$ coupling equal to 0.88 is taken from~\cite{DKKK}.

In Fig.~\ref{fig1} the moduli of the form factors $\Gamma^{s}_2(s)$ and  $\Gamma^{n}_2(s)$, and the transition function $G_1(s)$ divided by the constant $\chi$  are shown.
As seen in Eq.~(\ref{A2}) this division is necessary since both $\chi$ and $G_1(s)$ have dimension GeV$^{-1}$ while the kaon form factors are dimensionless.
  
The $P$-wave amplitude $A_3^-$ depends on the charged kaon decay constant $f_K$ and on the $B^-$ to $K^-$ vector transition form factor $F_1^{B^-K^-}(s_{23})$. 
Its form is taken from~\cite{Ball} as
\begin{equation}
\label{F1BK}
F_1^{BK}(s)=\frac{r_1}{1-\frac{s}{m_1^2}}+\frac{r_2}{(1-\frac{s}{m_1^2})^2}, 
\end{equation}
where $r_1=0.162$, $r_2=0.173$ and $m_1=5.41 $ GeV.

In Eq.~(\ref{A3}) there are two terms with the functions $F_u^{K^+K^-}(s_{23})$ and $F_s^{K^+K^-}(s_{23})$. The first term describes the transitions from the $u \bar{u}$ pair of quarks into a set of $\rho$ and $\omega$ resonances, while the second - the $s \bar{s}$ transitions into the $\phi(1020)$ and the $\phi(1680)$ resonances. 
Among the $\rho$ mesons we include three states: $\rho(770)$, $\rho(1450)$ and $\rho(1700)$, labelled by $i=1,2,3$, respectively.
In addition we add three isospin zero resonances: $\omega(782)$, $\omega(1420)$ and $\omega(1650)$ indicated by $i=4,5,6$. Thus the $F_u^{K^+K^-}(s_{23})$ function is parametrized as follows:
\be
\label{Fu}
F_u^{K^+K^-}(s_{23})=\sum_{i=1}^{6} b_i\, BW_i(s_{23}).
\ee
where $b_i$ are complex constants to be fitted to experimental data.
The Breit-Wigner functions read:
\be
\label{BWi}
BW_i(s_{23})=\frac{m_i^2}{m_i^2-s_{23} - i \sqrt{s_{23}}\, \Gamma_i(s_{23})}.
\ee
Here $m_i$ are masses of the  six vector mesons mentioned above and $\Gamma_i(s_{23})$ are their energy dependent widths
\be  
\label{Gammai}
\Gamma_i(s_{23})=\frac{m_i^2}{s_{23}}\, \Big(\frac{p}{p_i}\Big)^3\,\Gamma_i
\ee
with $\Gamma_i$ being the total widths of resonances and the momenta $p$ and $p_i$ defined as
\be
\label{ppi}
p=\frac{1}{2} \sqrt{s_{23}- 4 m_K^2},\,\,\,\,\,\,\,\,\,
p_i=\frac{1}{2} \sqrt{m_i^2- 4 m_K^2}.
\ee
Using the isospin symmetry for the $\rho$ and $\omega$ resonances we put $b_4=b_1$, $b_5=b_2$ and $b_6=b_3$ thus reducing the number of model parameters.

The $F_s^{K^+K^-}(s_{23})$ function is defined in a similar way as $F_u^{K^+K^-}(s_{23})$ in Eq.~(\ref{Fu}):
\be 
\label{Fs}
F_s^{K^+K^-}(s_{23})= b_{\phi} BW_{\phi}(s_{23}) +b_{\phi^{'}} BW_{\phi^{'}}(s_{23}).
\ee
Here the indices $\phi$ and $\phi^{'}$ label the resonances $\phi(1020)$ and $\phi(1680)$, respectively.
We treat $b_{\phi}$ as a real fitted parameter and take $b_{\phi^{'}}=0.018$ using
Table 2 of Ref.~\cite{Bruch}.

The term corresponding to a transition 
from the $d \bar{d}$ quarks into the final $K^+K^-$ pair and present in Eq.~(3) of Refs.~\cite{FKLZ} and ~\cite{LZ14} has been omitted since the   
two charged kaons have no $d \bar{d}$ constituents and therefore the $d \bar{d} \to K^+ K^-$ transition should be suppressed with respect to the $u \bar{u} \to K^+ K^-$ transition.

The amplitude $A_4^-$ represents the $B^-$ transitions into the $K^+K^-$ pairs in the $P$-wave. 
Therefore it is proportional to the vector transition form factor $A_0^{B^-\rho^0}(m^2_K)$ divided by the $\rho^0$ decay constant $f_{\rho}$ which in our approximation effectively represents contributions of all the $P$-wave resonances created from the $u \bar{u}$ pairs.
From Ref.~\cite{rhonumber} we take $A_0^{B^-\rho^0}(m^2_K)=0.37$. 

The amplitudes $A_5^-$ and $A_6^-$ depend on the function $D(s_{12},s_{23})$
which is expressed as
\begin{equation}
\label{De}
D(s_{12},s_{23}) 
=\frac{1}{3}(\vert \overrightarrow{p_1} \vert \vert \overrightarrow{p_2} \vert )^2
-(\overrightarrow{p_1}\cdot \overrightarrow{p_2})^2,
 \end{equation}
where 
$\overrightarrow{p_1}$ and $\overrightarrow{p_2}$  are the momenta of the kaons $K^-(p_1)$ and the $K^+(p_2)$ in the rest frame of $K^+(p_2)$ and $K^-(p_3)$.
If one denotes $m_{23}=\sqrt{s_{23}}$ then 
\begin{eqnarray}
\label{kincms23}
\overrightarrow{p_1}\cdot \overrightarrow{p_2}&=&\frac{1}{4}(s_{13}-s_{12}), \nonumber \\
\vert \overrightarrow{p_1} \vert  & = &  \frac{1}{2m_{23}}\sqrt{\left[M_B^2-\left(m_{23}+m_K\right)^2\right]
\left[M_B^2-\left(m_{23}-m_K\right)^2\right]},
\nonumber \\
\vert
 \overrightarrow{p_2} \vert & = & \frac{1}{2}\sqrt{m_{23}^2-4m_K^2}. 
\end{eqnarray}

Using the mixing angle $\alpha_T=(81\pm 1)^{\circ}$ relevant for the quark composition of $f_2(1270)$ and $f_2^{'}(1525)$ (see Ref.~\cite{PDG22}) one gets
\begin{equation}
\langle f_2|u\bar{u}\rangle = \frac{1}{\sqrt{2}}\sin \alpha_T \approx 0.698
\end{equation}
and
\begin{equation}
\langle f_2^{'}|u\bar{u}\rangle = \frac{1}{\sqrt{2}}\cos \alpha_T \approx 0.111.
\end{equation}
These factors are included in Eqs.~(\ref{A5}) and (\ref{A6}), respectively.

The amplitude $A_5^-$ is proportional to the Breit-Wigner function $G_{f_2K^+K^-}(s_{23})$ describing a coupling of the $f_2(1270)$ resonance to the final $K^+K^-$ pair:
\begin{equation}
\label{G2f2}
G_{f_2K^+K^-}(s_{23})=\frac{g_{f_2K^+K^-}}{m^2_{f_2}-s_{23}-im_{f_2}\Gamma_{f_2}}.
\end{equation}
The coupling constant $g_{f_2K^+K^-}$ is evaluated from the expression
\begin{equation}
\label{gf2}
g_{f_2K^+K^-}=m_{f_2}\sqrt{\frac{60 \pi \Gamma_{f_2K^+K^-}}{q^5_{f_2}}},
\end{equation}
where $q_{f_2}$ is the kaon momentum in the $f_2$ center-of-mass frame and $\Gamma_{f_2K^+K^-}=\frac{1}{2}\cdot 4.6\% \cdot \Gamma_{f_2}$, with $m_{f_2}$ and $\Gamma_{f_2}$ being the $f_2(1270)$ mass and its total width taken from Ref. \cite{PDG22}.
The last factor in Eq.~(\ref{A5}) is the effective form factor $F^{Bf_2}(m^2_K)$ for the transition of the $B^-$ meson into $f_2(1270)$. It will be treated as a free complex parameter called further $F_5$.\footnote{The effective form factor for the $B \to f_2$ transition is composed of three terms which are not well known (see Eq. (10a) in Ref. \cite{Koreans}).
Here we have to correct the values of $F^{Bf_2}(m^2_K)$ given in Table 1 of Ref.~\cite{LZ14} for the two fits to data as they involve numerical errors. They should be rescaled down:
the value $11.9\pm 1.3$ should be replaced by $1.08\pm 0.12$, while the value $12.3\pm 3.5$
 by $1.11 \pm 0.32$. 
} 

The Breit-Wigner function $G_{{f_2} {^{'}} K^+K^-}(s_{23})$ in the $A_6^-$ amplitude from Eq.~(\ref{A6}) is defined in a similar way as the corresponding
 $G_{f_2 K^+K^-}(s_{23})$ function in Eqs.~(\ref{G2f2}) and (\ref{gf2}).
The width for the $f_2^{'}(1525)$ decay into $K^+K^-$ is equal to $\Gamma_{f_2^{'}K^+K^-}=\frac{1}{2}\cdot 88.7\% \cdot \Gamma_{f_2^{'}}$.
The effective form factor $F^{B^{-}{f_2}'}(m_K^2)$, hereafter named $F_6$, will be 
a complex parameter to be determined from a fit to data. 

The $B^{+} \to K^{+}K^-K^+$ decay amplitude can be obtained from the $A^-_{sym}$ amplitude written in Eq. (\ref{A-sym}) by substitutions $\Lambda_u \to \Lambda_u^*$ and $\Lambda_c \to \Lambda_c^*$ in the definitions of the coefficients $v, y, w_u$ and $w_s$ appearing in Eqs. (\ref{A1} - \ref{A6}).
Moreover, one has to change $B^-$ into $B^+$ and $K^-$ into $K^+$ in the corresponding equations.\\

\section{Dalitz plot distributions} 
\label{Sect3}

The double differential branching fraction distributions which can also be called the Dalitz plot distributions for the $B^{\pm} \to K^{\pm}(p_1) K^+(p_2) K^-(p_3)$ decays are expressed by the decay amplitudes $A^{\pm}_{sym}(s_{12},s_{23})$ as:

\be
\label{dBr}
\frac{dBr^{\pm}}{ds_{12}\, ds_{23}}=\frac{1}{32 (2 \pi)^3 M_B^3 \Gamma_B} |A^{\pm}_{sym}(s_{12},s_{23})|^2,
\ee
where $\Gamma_B$ is the total width of the charged $B$ meson.

As the Dalitz plot distribution is symmetric under the exchange of $s_{12}$ and $s_{23}$,  one can limit the integration range on $s_{12}$ to the values larger than $s_{23}$.
Taking this into account we introduce new names of the relevant variables, ie. $m_{K^+K^-\,low}=m_{23}=\sqrt{s_{23}}$ and $m_{K^+K^-\,high}=m_{12}=\sqrt{s_{12}}$.
An additional factor of 2 is inserted in
the differential effective $m_{23}$ mass distribution:
\be
\label{dBrlow}
\frac{dBr^{\pm}}{d m_{23}}=4\, m_{23}\int _{D_{12}}^{s_{12\, max}} d s_{12}\,
\frac{dBr^{\pm}}{ds_{12}\, ds_{23}}.
\ee
Here $D_{12}$ is equal to $s_{12\,min}$ for $2 m_K <m_{23}<d$ or $D_{12}=s_{23}$
for $d < m_{23} < M_B - m_K$, where $d=\sqrt{m_K (M_B+m_K)}\approx 1.6882$ GeV.
The limits on the values of $s_{12}$ at the border of the Dalitz plot contour for fixed
values of $m_{23}$ can be calculated in the center-of-mass frame
of the $K^{\pm}(p_2)K^{\mp}(p_3)$ mesons:
\begin{eqnarray}
\label{s12lim}
s_{12\,min}=(E_1+E_2)^2-(\vert \overrightarrow{p_1} \vert +\vert \overrightarrow{p_2} \vert )^2,
\\ \nonumber
s_{12\,max}=(E_1+E_2)^2-(\vert \overrightarrow{p_1} \vert -\vert \overrightarrow{p_2} \vert)^2, 
\end{eqnarray}
where
\begin{eqnarray}
\label{E12lim}
E_1=\frac{M_B^2-m_{23}^2-m_K^2}{2m_{23}},\,\,\,\,\,\,\,\,\vert \overrightarrow{p_1} \vert =\sqrt{E_1^2-m_K^2},
\\ \nonumber
E_2=\frac{m_{23}}{2},\,\,\,\,\,\,\,\, \vert \overrightarrow{p_2} \vert =\sqrt{E_2^2-m_K^2}.
\end{eqnarray}
The lower limit of $m_{23}$ is equal to the $K^+K^-$ threshold value $2\,m_K\approx
0.98735$ GeV and the upper limit is $g=\sqrt{(M_B^2-m_K^2)/2}\approx 3.7166$ GeV.

The second branching fraction projection or the differential effective $m_{12}$ mass distribution is given by the following expression:
\be
\label{dBrhigh}
\frac{dBr^{\pm}}{d m_{12}}=4\, m_{12} \int^{G_{23}}_{s_{23\, min}} d s_{23}\,
\frac{dBr^{\pm}}{ds_{12}\, ds_{23}}.
\ee
In this equation $m_{12}$ varies between the values $d$ and $M_B-m_K$ and $G_{23}$
is equal to $m_{12}^2$ for $d<m_{12}<g$ or it is equal to $s_{23\,max}$ for
$g<m_{12}<M_B-m_K$. 
The limits of the $s_{23}$ values at fixed
$m_{12}$ can be calculated in the center-of-mass frame
of the $K^{\mp}(p_1)K^{\pm}(p_2)$ mesons:
\begin{eqnarray}
\label{s23lim}
s_{23\,min}=(E_2^{'}+E_3)^2-( \vert \overrightarrow{p_2^{\,\,,}} \vert 
+ \vert \overrightarrow{p_3} \vert )^2,
\\ \nonumber
s_{23\,max}=(E_2^{'}+E_3)^2-( \vert \overrightarrow{p_2^{\,\,,}} \vert 
- \vert \overrightarrow{p_3} \vert )^2, 
\end{eqnarray}
where
\begin{eqnarray}
\label{E23lim}
E_2^{'}=\frac{m_{12}}{2},\,\,\,\,\,\,\,\,\,\,\,\,
\vert \overrightarrow{p_2^{\,\,,}} \vert =\sqrt{E_2^{'2}-m_K^2},
\\ \nonumber
E_3=\frac{M_B^2-m_{12}^2-m_K^2}{2m_{12}},\,\,\,\,\,\,\,\,\,\,\,\,\,\,\,
\vert \overrightarrow{p_3} \vert = \sqrt{E_3^2-m_K^2}. 
\end{eqnarray}

There is also a possibility to make the third Dalitz plot projection as the differential effective $m_{13}$ mass distribution. In the case of the $B^{-} \to K^{-} K^+ K^-$ decay this is the $m_{K^-K^-}$ distribution and in the case of the $B^{+} \to K^{+} K^+ K^-$ reaction we have the $m_{K^+K^+}$ distribution.
The integration formula for these distributions reads:
\be
\label{dBr13}
\frac{dBr^{\pm}}{d m_{13}}=4\, m_{13} \int^y_x d s_{23}\,
\frac{dBr^{\pm}(s_{12}= 2y-s_{23},s_{23})}{ds_{12}\, ds_{23}},
\ee
where $y=(M_B^2+3 m_K^2-s_{13})/2$, $s_{13}=m_{13}^2$ and 
$x=y-\sqrt{y^2-m_K^2(M_B^2-m_K^2)^2/s_{13}}$.
Here the equality $s_{12}+s_{13}+s_{23}=M_B^2+3 m_K^2$ has been applied.

In Ref.~\cite{LHCb14} the LHCb Collaboration has presented the $m_{K^+K^-\,low}$ distributions for the signal events splitted according to the sign of
$cos\,\theta_H$, where $\theta_H$ is the helicity angle. 
For the $B^{\mp} \to K^{\mp}(p_1)K^{\pm}(p_2)K^{\mp}(p_3)$ decay $\theta_H$ is defined in the center-of-mass frame of the $K^{\pm}(p_2)K^{\mp}(p_3)$ mesons as the angle between the two same-sign charge kaons.
The variables $s_{12}$ and $s_{23}$ are related to $cos\,\theta_H$ in the following way:
\be 
\label{m12}
s_{12}=\frac{1}{2}(M_B^2-s_{23}+3 m_K^2)+2 \vert \overrightarrow{p_1} \vert
\vert \overrightarrow{p_2} \vert cos\,\theta_H ,
\ee
where the moduli of the momenta $\vert \overrightarrow{p_1} \vert$ and
$\vert \overrightarrow{p_2} \vert$ are given in Eq.~(\ref{E12lim}).
Then from Eqs.~(\ref{s12lim}) one can see that the lower limit of $s_{12}$ corresponds
to $cos\,\theta_H=-1$ while $s_{12\,max}$ corresponds to $cos\,\theta_H=+1$.
\footnote{We have noticed a disagreement between the definition of the helicity angle
and the captions of Figs.\,4,\,5 and 6 given in Ref.~\cite{LHCb14}, namely the figures corresponding to the regions 
of cos$\,\,\theta_H>0$ and cos$\,\,\theta_H<0$ have been interchanged. Therefore (for example)  the third line of Fig.\,6 caption should read:
"The plots are restricted to events with (a),(c) cos$\,\,\theta > 0$ and (b), (d) cos$\,\,\theta < 0$." The necessity of this change has been confirmed by Irina Nasteva (Ref.~\cite{Irina}).}

Using the above definitions of $s_{12}$ limits we are able to split the $m_{23}$ distribution from Eq.~(\ref{dBrlow}) into two parts corresponding to $cos\,\theta_H <0$ 
\be 
\label{dBr<0}
\frac{dBr^{\pm}(cos\,\theta_H <0)}{d m_{23}}=4\, m_{23}\int _{DL}^{GL} d s_{12}\,
\frac{dBr^{\pm}}{ds_{12}\, ds_{23}}
\ee 
or to $cos\,\theta_H >0$ 
\be 
\label{dBr>0}
\frac{dBr^{\pm}(cos\,\theta_H >0)}{d m_{23}}=4\, m_{23}\int _{DG}^{s_{12\, max}} d s_{12}\,
\frac{dBr^{\pm}}{ds_{12}\, ds_{23}}.
\ee 
The lower limit $DL$ of the first integral equals to $s_{12\,min}$ for $2 m_K<m_{23}<d$ and $DL=s_{23}$ for
$d<m_{23}<m_0\equiv \sqrt{M_B^2/3+m_K^2}\approx 3.0877$ GeV.
The upper limit is given by $GL=(M_B^2-s_{23}+3 m_K^2)/2$.
Finally the lower limit $DG$ in Eq.~(\ref{dBr>0}) equals to $GL$ for $2 m_K<m_{23}<m_0$ or $DG=s_{23}$ if $m_0<m_{23}<g$.

\section{Data selection, the fitting method and additional parameters in the model }
\label{data}
Three data sets of the $B^{\pm} \to K^{\pm}K^+K^-$ decays measured by the $BABAR$ 
(Ref.~\cite{BABAR12}) and by the LHCb collaborations (Refs.~\cite{LHCb14}, 
~\cite{LHCb16} and ~\cite{LHCb23}) are simultaneously analysed.
The effective $K \bar{K}$ mass distributions $m_{K^+K^-~low}$, $m_{K^+K^-~high}$ and $m_{K^+K^+}$ presented by $BABAR$ on Figs.~7 and 8 of \cite{BABAR12} and 
the LHCb projections from Fig.~6 of \cite{LHCb14}, from Fig.~2 of \cite{LHCb16} and from Fig.~7(a) of \cite{LHCb23} are chosen for a comparison with the theoretical calculations based on the model described in Secs.~\ref{Amplitudes} and~\ref{Sect3}.
The model parameters have also been constrained by the experimental branching fraction of the $B^+ \to K^+K^-K^+$ decay: $Br_{exp} = (3.40 \pm 0.14) \times
10^{-5}$ \cite{PDG22}.
Another constraint has been related to the experimental value of the branching fraction for the decay of $B^+$ into $K^+\phi$ multiplied by the secondary branching fraction for the decay of $\phi$ into the $K^+K^-$ pair (Ref.~\cite{PDG22}).
The resulting value for the $B^+ \to K^+K^-K^+$ decay in the range of the $m_{K+K^-\,low}$ mass where the $P$-wave $\phi(1020)$ resonance dominates has been taken as
$Br_{exp}^P = (4.33 \pm 0.34) \times 10^{-6}$. 
This value has been compared to the integral of the theoretical branching fraction distribution $\frac{dBr^{+}}{d m_{23}}$ over the mass range between the $K^+K^-$ threshold and 1.05 GeV.
Hovewer, in this case we have only included the $P$-wave amplitudes $A_3^+$ and $A_4^+$ in the amplitude $A^+_{sym}(s_{12},s_{23})$ (see Eq.~\ref{dBr}).

In order to make a comparison of the experimental Dalitz plot projections with our model
we have eliminated a few experimental data points corresponding to the $K \bar{K}$ masses in bins with the central values below the nominal $K^+K^-$ threshold mass or the masses which can be attributed to the decays of the $D^0$, $J/\Psi$, $\chi_{c0}$ and $\Psi(4360)$ mesons to $K^+K^-$. Here we note that the above decays are not included in the present model.

The theoretical branching fraction distributions defined in Sec.~\ref{Sect3} have been multiplied by the normalization factors defined as the ratios of the $B^{\pm}$ signal yields to the total branching fraction $Br_{exp}$.
For the $BABAR$ data~\cite{BABAR12} the number of signal events is equal to 
5269.
According to the LHCb data from Figs.~2(a) and 2(b) of Ref.~\cite{LHCb16} a sum of the $B^+$ and $B^-$ signal events is 103211 and this number is used to calculate the corresponding normalization factor.
The normalization factor of the LHCb data from Fig. 7(a) in \cite{LHCb23} is calculated as a ratio of the sum of the $B^+$ and $B^-$ signal events to the theoretical branching fraction integrated over the $m^2(K^+K^-)_{high}$ distribution while the 
$m^2(K^+K^-)_{low}$ region is limited by the 1.1 GeV$^2$ and 2.25 GeV$^2$ values.
One bin corresponding to $J/\Psi$ and three bins around the $\chi_{c0}$ position have been subtracted from the data set.

Altogether 318 $BABAR$ data bins, 256 LHCb bins from ~\cite{LHCb14} and 
~\cite{LHCb16}, and 142 bins from~\cite{LHCb23} have been taken into account, so the total number of the data bins was equal to 716.
The $\chi^2$ distribution method has been used to fit all these data and in addition the
two total branching fractions $Br_{exp}$ and $Br_{exp}^P$. In order to get a better
normalization of the model amplitudes to the full data set the $\chi^2$ values for
these two last data have been multiplied by a factor of ten .

After performing some preliminary fits to data we realized that the initial set of 17 model parameters is not sufficient to describe well the Dalitz plot projection distributions, especially for the high effective $K\bar{K}$ masses above 1.8 GeV. 
This is a region where our knowledge of the kaon-kaon interactions is very limited,
in particular in the $S$-wave.
Thus we have modified the functional dependence of two $S$-wave amplitudes $A_1^-$ and $A_2^-$ on the $s_{23}$ variable.
The kaon strange scalar form factor $\Gamma^{s*}_2(s_{23})$ in Eq.~(\ref{A1})
has been multiplied by the following polynomial
\be 
\label{Ps}
P_s(x)=(1+\sum_{i=1}^{6} c_i x^i)f_1,\,\,\,\,\,\,\,\,x=m_{23}-2 m_K,
\ee
where $c_i$ are real coefficients, $f_1$ is complex and we remind that $m_{23}=\sqrt{s_{23}}$.
This polynomial is normalized to 1 at the $K^+K^-$ threshold.
Similarly the kaon non-strange scalar form factor $\Gamma^{n*}_2(s_{23})$ in Eq.~(\ref{A2})
has been multiplied by the polynomial $P_n(x)$:
\be 
\label{Pn}
P_n(x)=1+\sum_{i=1}^{6} d_i x^i,
\ee
where $d_i$ are also real coefficients.
Finally we multiply the transition function $G_1(s_{23})$ by the polynomial $P_G(x)$
\be 
\label{PG}
P_G(x)=1+\sum_{i=1}^{6} g_i x^i,
\ee
where $g_i$ are new real parameters. 

The same polynomials $P_s(x)$, $P_n(x)$ and $P_G(x)$ have been introduced in the $B^+$ decay amplitude $A^+(s_{12},s_{23})$.
Addition of these polynomials in the $S$-wave amplitudes has led us to a substantial improvement of the theoretical  distributions on the Dalitz plot.
The total number of free parameters equals to 35.\\

\section{Results}
\label{results}

The parameters obtained in the data fit are given in Table~\ref{parameters} 
and the coefficients of the polynomials defined in Eqs.~(\ref{Ps}-\ref{PG}) can be found in Table~\ref{polynomials}.
The theoretical value of the total averaged branching fraction for the $B^{\pm} \to K^{\pm}K^-K^+$ decays equal to $Br^{th}=3.33\times 10^{-5}$ is very close to the experimental branching fraction $Br^{exp}=(3.40 \pm 0.14) \times10^{-5}$.
Also the total branching fraction integrated over the $K^+K^-$ effective mass up to 1.05 GeV, where the $\phi(1020)$ resonance dominates, has been calculated. Its value $4.45\times 10^{-6}$ lies well within one standard deviation from the experimental value $(4.33 \pm 0.34) \times 10^{-6}$.

\begin{table}
\caption{Parameters of our model amplitudes. Phases are given in radians.}
\begin{center}
\begin{tabular}{cccc}
\hline
\hline
Amplitude & Parameter& modulus & phase  \\ 
          parts &  \vspace*{0.2cm}\\
\hline
$A_1$& $f_1$  &   0.60371  & -0.0504 \vspace*{0.2cm} \\   
$A_2$& $\chi$ & 1.4729 GeV$^{-1}$ & -1.4484 \vspace*{0.2cm} \\ 
$A_2$& $r_2$  & 11.166 GeV$^{3/2}$ &-2.4061 \vspace*{0.2cm} \\ 
$A_3, A_4$ & $b_1$  &  4.0331 &  2.8885 \vspace*{0.1cm} \\
$A_3, A_4$ & $b_2$  &  0.19908 &  -2.0 \vspace*{0.1cm} \\
$A_3, A_4$ & $b_3$  &  0.14769&  0.5230 \vspace*{0.1cm} \\
$A_3$  & $b_{\phi}$ &1.1524& 3.1416  \vspace*{0.1cm} \\
$A_5$& $F_5$        &0.15961 &  0.0793 \vspace*{0.1cm} \\
$A_6$&  $F_6$       &1.0192 &  2.9281 \vspace*{0.1cm} \\ 
\hline
\hline
\end{tabular} 
\end{center}
\label{parameters}
\end{table}
 
\begin{table}
\caption{Coefficients of polynomials in our model amplitudes
(Eqs.~\ref{Ps}\,-~\ref{PG}). Dimensions of the coefficients are GeV$^
{-i}$, where the index $i=1,...\,,6$.}
\begin{center}
\begin{tabular}{cccc}
\hline
\hline
Index & polynomial $P_s(x)$ &  polynomial $P_n(x)$&polynomial $P_G(x)$\vspace*{0.1cm} \\ 
$i$  &  $c_i$  &  $d_i$  &  $g_i$
\vspace*{0.2cm} \\ 
\hline
1 &  -0.57635 &  -7.2486 &   2.1473 \vspace*{0.1cm} \\   
2 &  -2.3976 &   -4.6746 &  -0.10438 \vspace*{0.1cm} \\ 
3 &   3.7440 &   32.407  &   1.4636 \vspace*{0.1cm} \\ 
4 &  -1.3362 &  -22.732  &   4.6728 \vspace*{0.1cm} \\
5 &  .052993  &   4.4602 &  -3.6713 \vspace*{0.1cm} \\
6 &  .017296 &   -0.16001 &  0.56857 \vspace*{0.1cm} \\ 
\hline
\hline
\end{tabular} 
\end{center}
\label{polynomials}
\end{table}

\begin{figure}
\begin{center}
 \includegraphics[width=8cm]{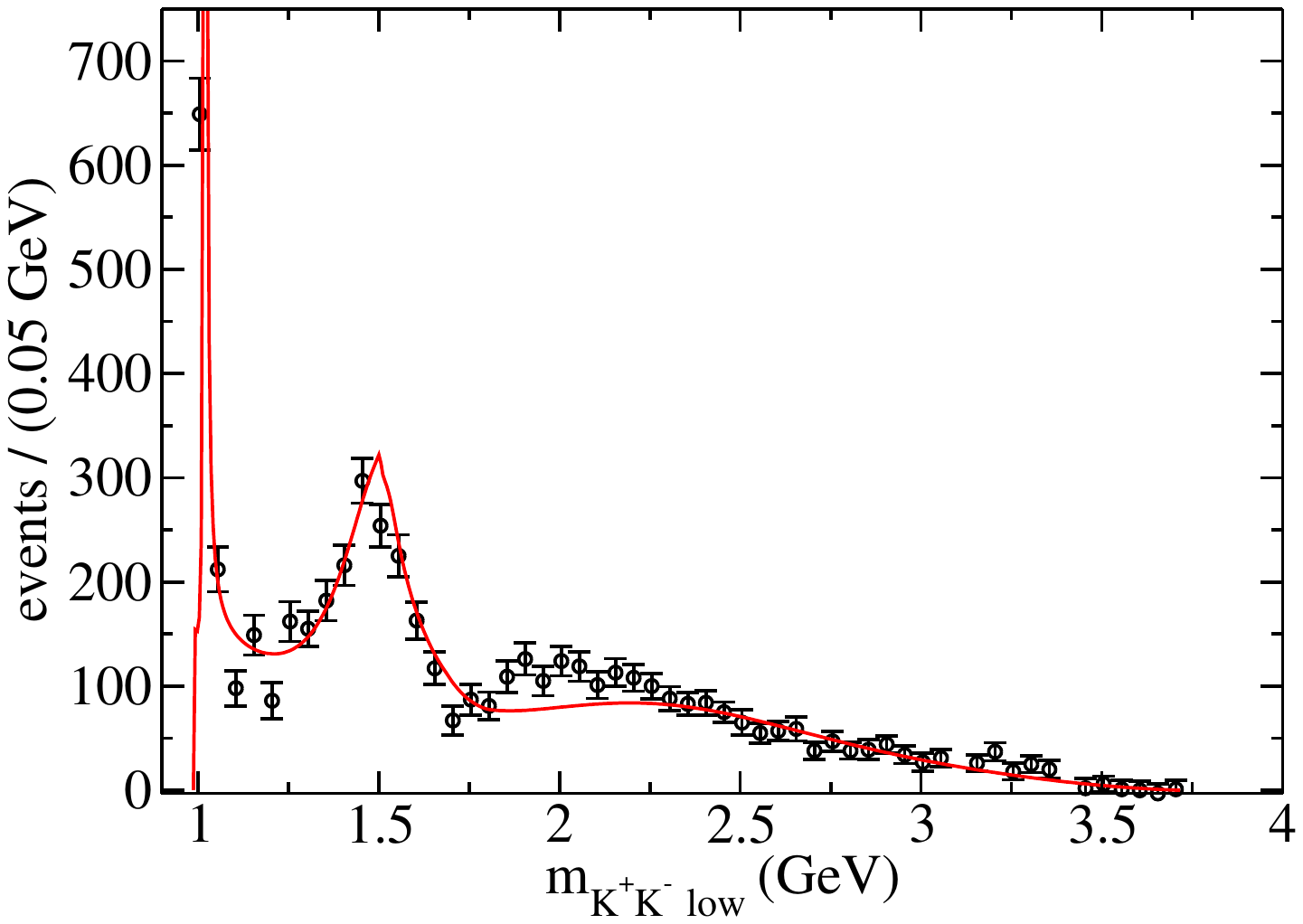}
\vspace*{6pt}
\caption{\label{fig2} Distribution of $m_{K^+K^-\,low}$. The $BABAR$ Collaboration data points are taken from Fig. 7 of Ref.~\cite{BABAR12}. The line is calculated from  our model by adding the $B^-$ and $B^+$ distributions given by Eq.~(\ref{dBrlow}).}
\end{center}
\end{figure}

\begin{figure}[h]
\begin{center}
\includegraphics[width=8cm]{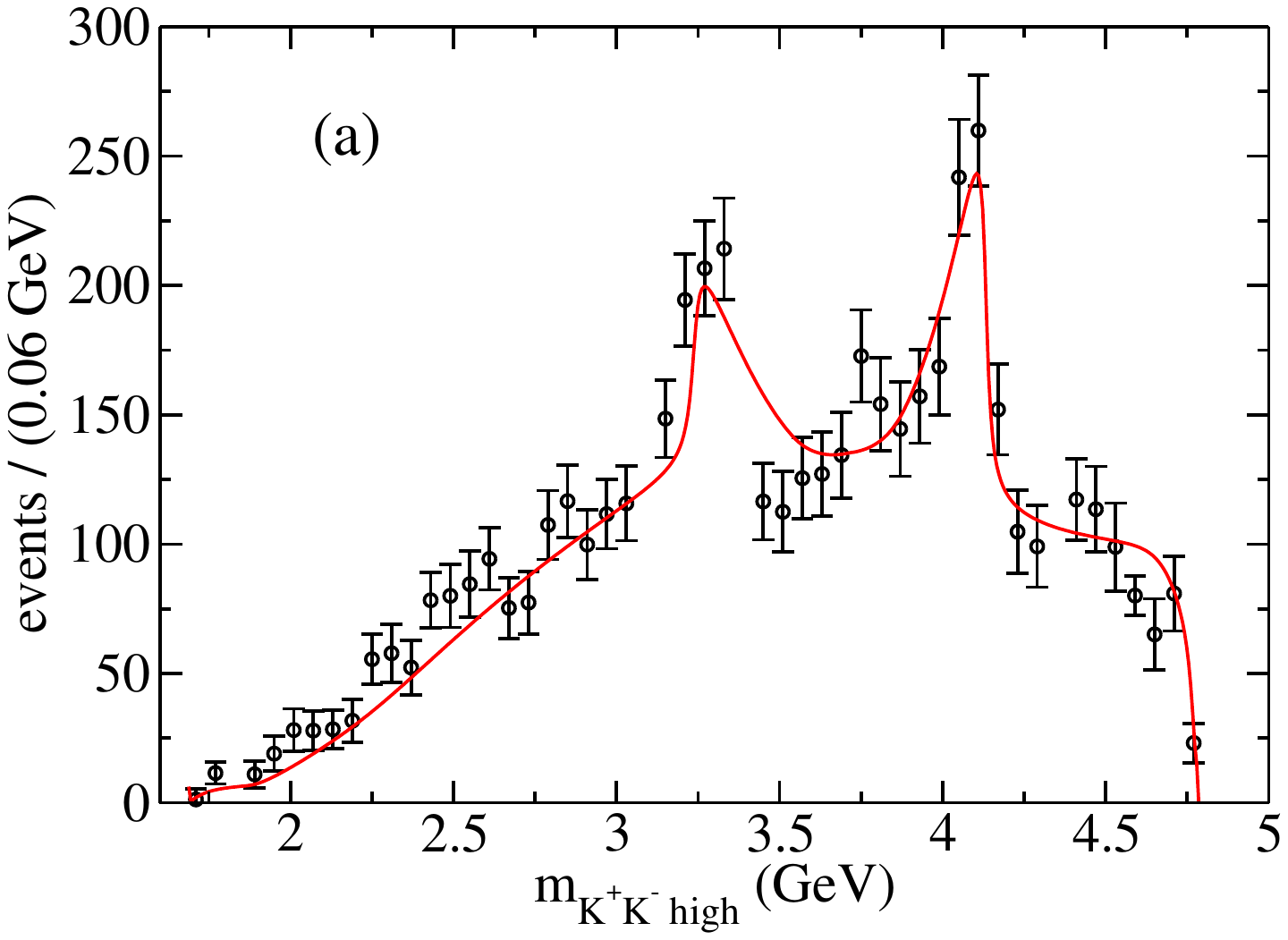}~~
\includegraphics[width=8cm]{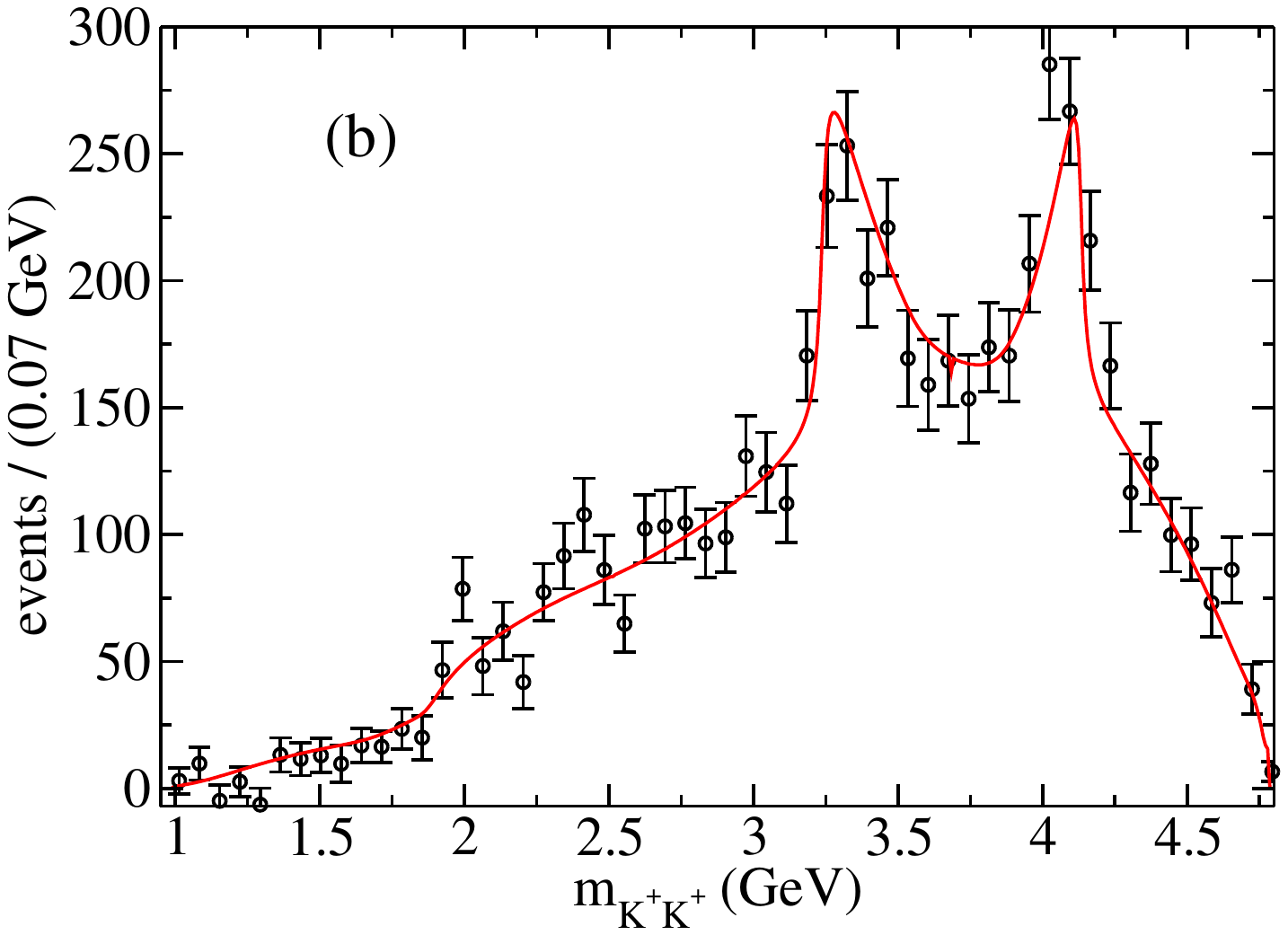}
\end{center}
\caption{\label{fig3} Distributions of $m_{K^+K^-\,high}$ (a) and 
$m_{K^+K^+}$ (b). The $BABAR$ Collaboration data points are taken from Fig.~7 of Ref.~\cite{BABAR12}. The lines are calculated from our model by adding the $B^-$ and $B^+$ distributions given by Eq.~(\ref{dBrhigh}) for the case (a) and by Eq.~(\ref{dBr13}) for the case (b).}
\end{figure}

\begin{figure}[h]
\begin{center}
\includegraphics[width=8cm]{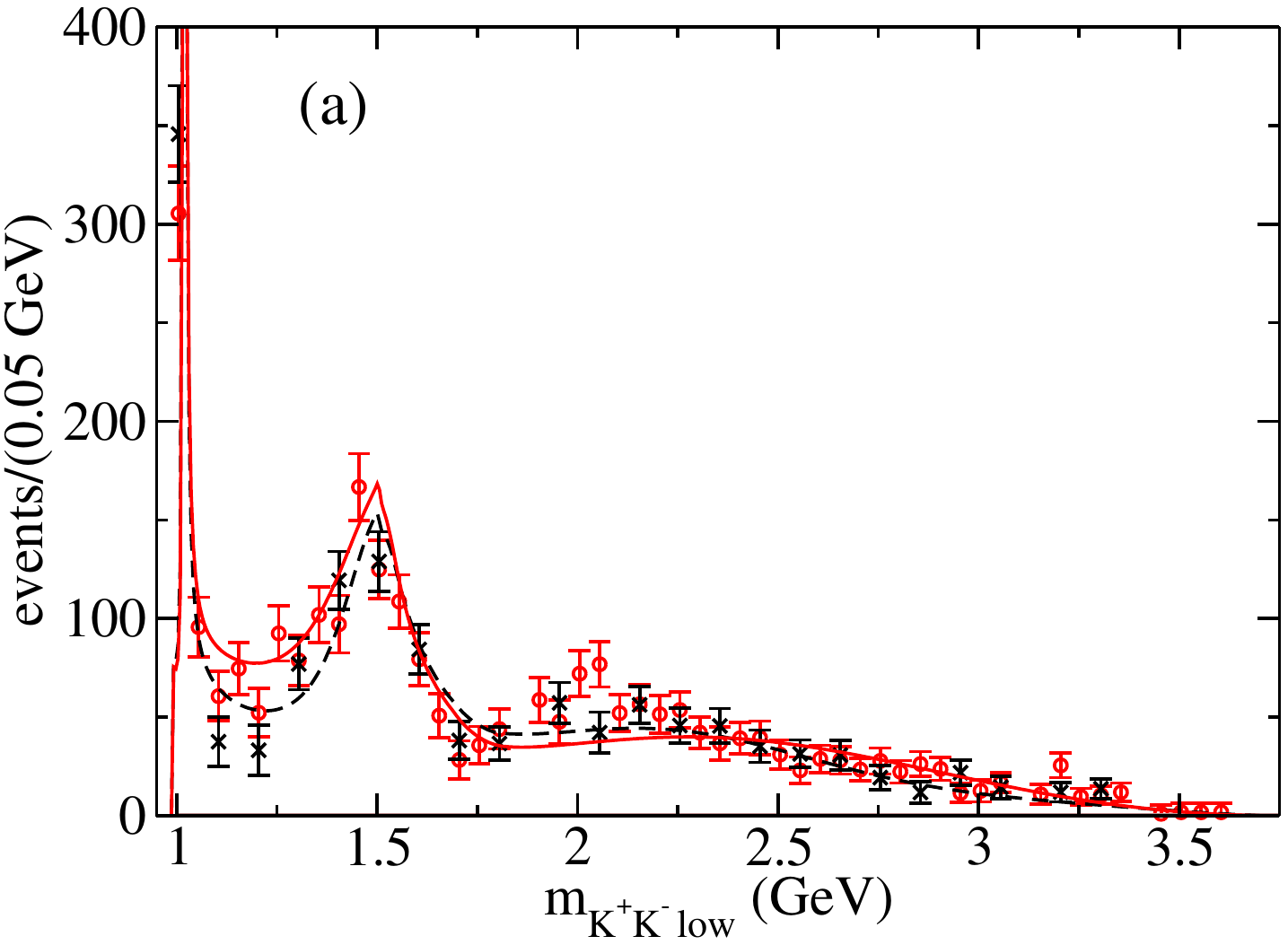}~~
\includegraphics[width=8cm]{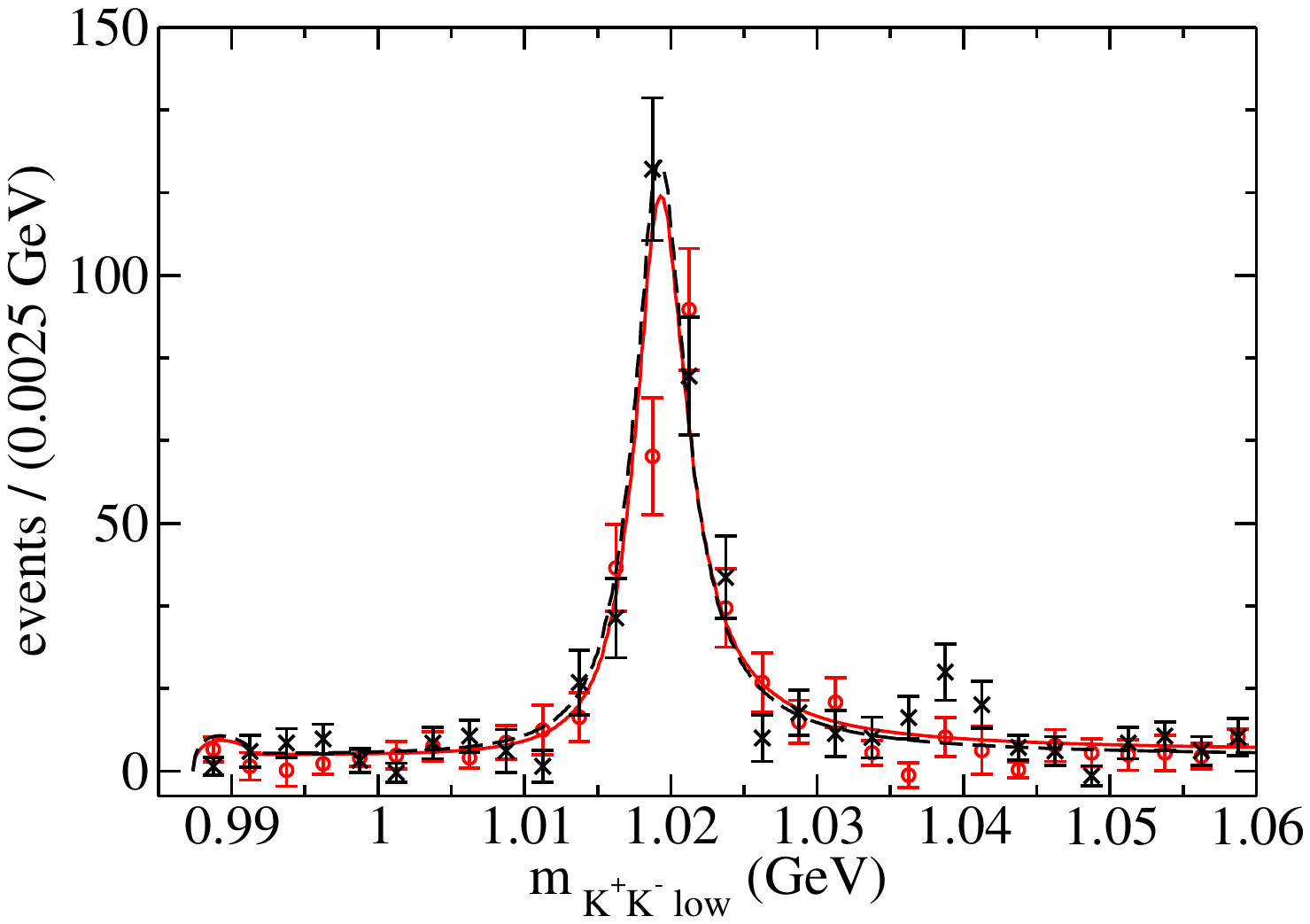}
\end{center}
\caption{\label{fig4}
Distribution of $m_{K^+K^-\,low}$ in the full mass region (a) and in the mass range restricted to 1.06 GeV (b). The $BABAR$ Collaboration data points for the 
$B^{\pm} \to K^{\pm}K^+K^-$ decays are taken from Fig.~8 of Ref.~\cite{BABAR12}. The red points correspond to the $B^+$ decays while the black ones to the $B^-$ events.
The red continuous lines represent the results of our model for the $B^+$ decays and the black dashed lines correspond to the $B^-$ decays.}
\end{figure}

\begin{figure}[h]
\begin{center}
\includegraphics[width=8cm]{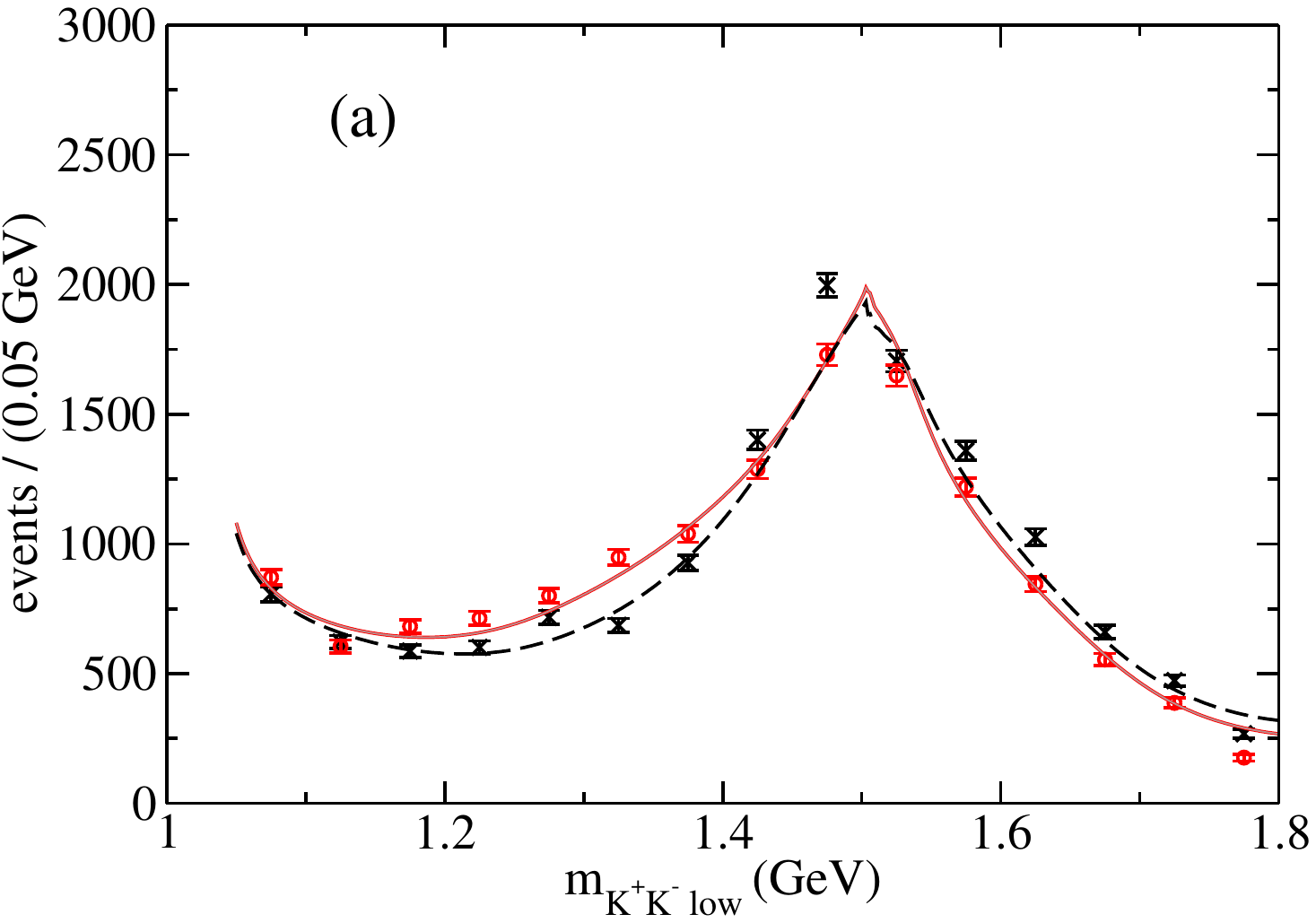}~~
\includegraphics[width=8cm]{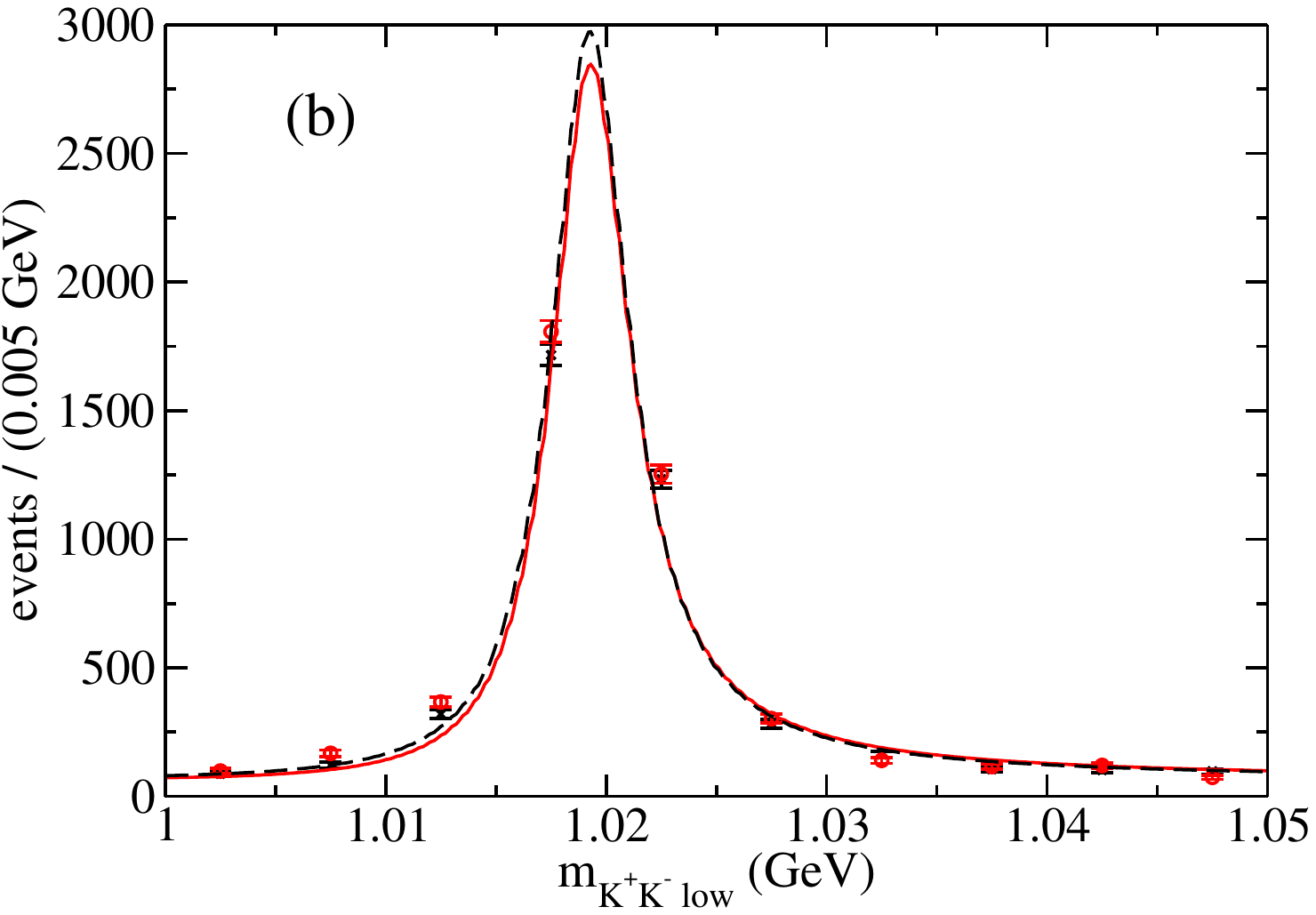}
\end{center}
\caption{\label{fig5}
Distributions of the $m_{K^+K^-\,low}$ variable in the mass region up to 1.8 GeV (a) and in the mass range dominated by the $\phi(1020)$ resonance (b).
The LHCb Collaboration data for the 
$B^{\pm} \to K^{\pm}K^+K^-$ decays are taken from Fig. 6(a) of Ref.~\cite{LHCb14}.
The events are chosen from the range $cos\,\theta_H >0$, where $\theta_H$ is the helicity angle.
The red points correspond to the $B^+$ decays while the black ones to the $B^-$ events.
The continuous red lines are calculated from our model for the $B^+$ decays while the
black dashed ones for the $B^-$ decays.
}
\end{figure}

The Dalitz plot projections obtained from our model and compared with the corresponding $BABAR$ data~\cite{BABAR12} are shown in Figs.\,\ref{fig2}\,-\,\ref{fig4}.
The theoretical distributions compared with the LHCb data sets~\cite{LHCb14},\,\cite{LHCb16} and  \cite{LHCb23} can be found in Figs.\,\ref{fig5}\,-\,\ref{fig8}.

\begin{figure}[h]
\begin{center}
\includegraphics[width=8cm]{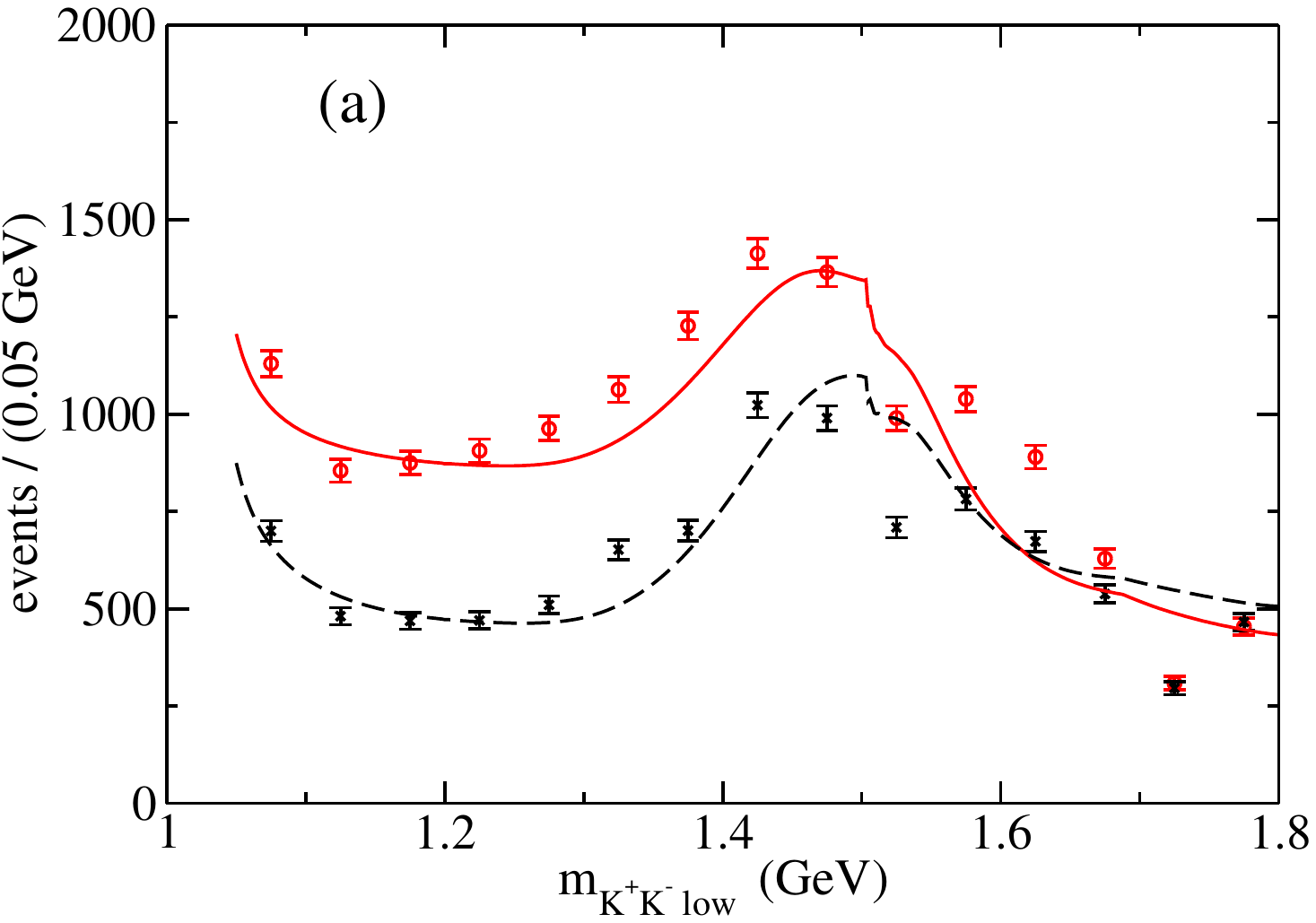}~~
\includegraphics[width=8cm]{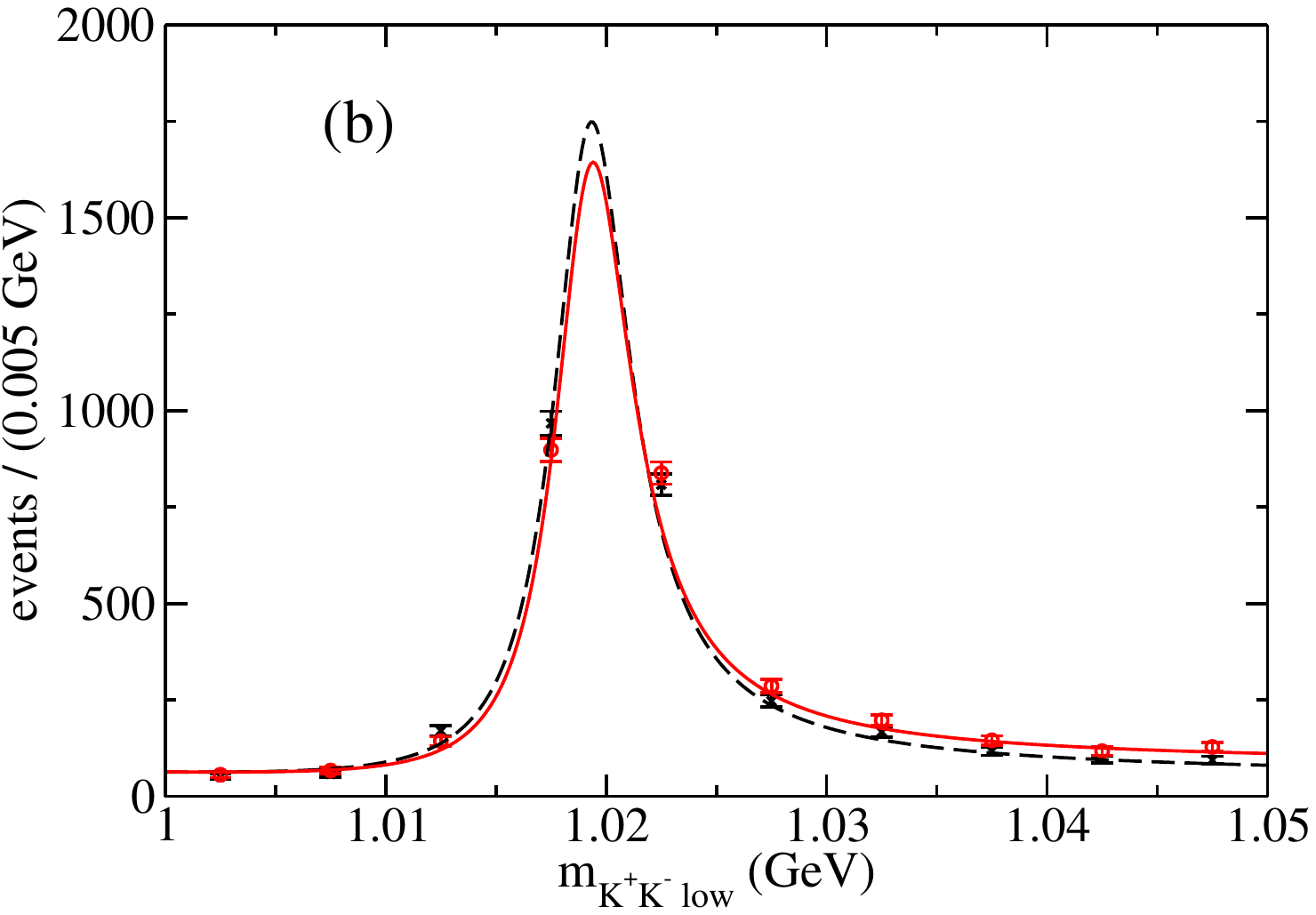}
\end{center}
\caption{\label{fig6}
Distributions of the $m_{K^+K^-\,low}$ variable in the mass region up to 1.8 GeV (a) and in the mass range dominated by the $\phi(1020)$ resonance (b).
The LHCb Collaboration data for the 
$B^{\pm} \to K^{\pm}K^+K^-$ decays are taken from Fig. 6(b) of Ref.~\cite{LHCb14}.
The events are chosen from the range $cos\,\theta_H <0$, where $\theta_H$ is the helicity angle.
The red points correspond to the $B^+$ decays while the black ones to the $B^-$ events.
The continuous red lines are calculated from our model for the $B^+$ decays while the
black dashed ones for the $B^-$ decays.
}
\end{figure}

\begin{figure}[h]
\begin{center}
\includegraphics[width=8cm]{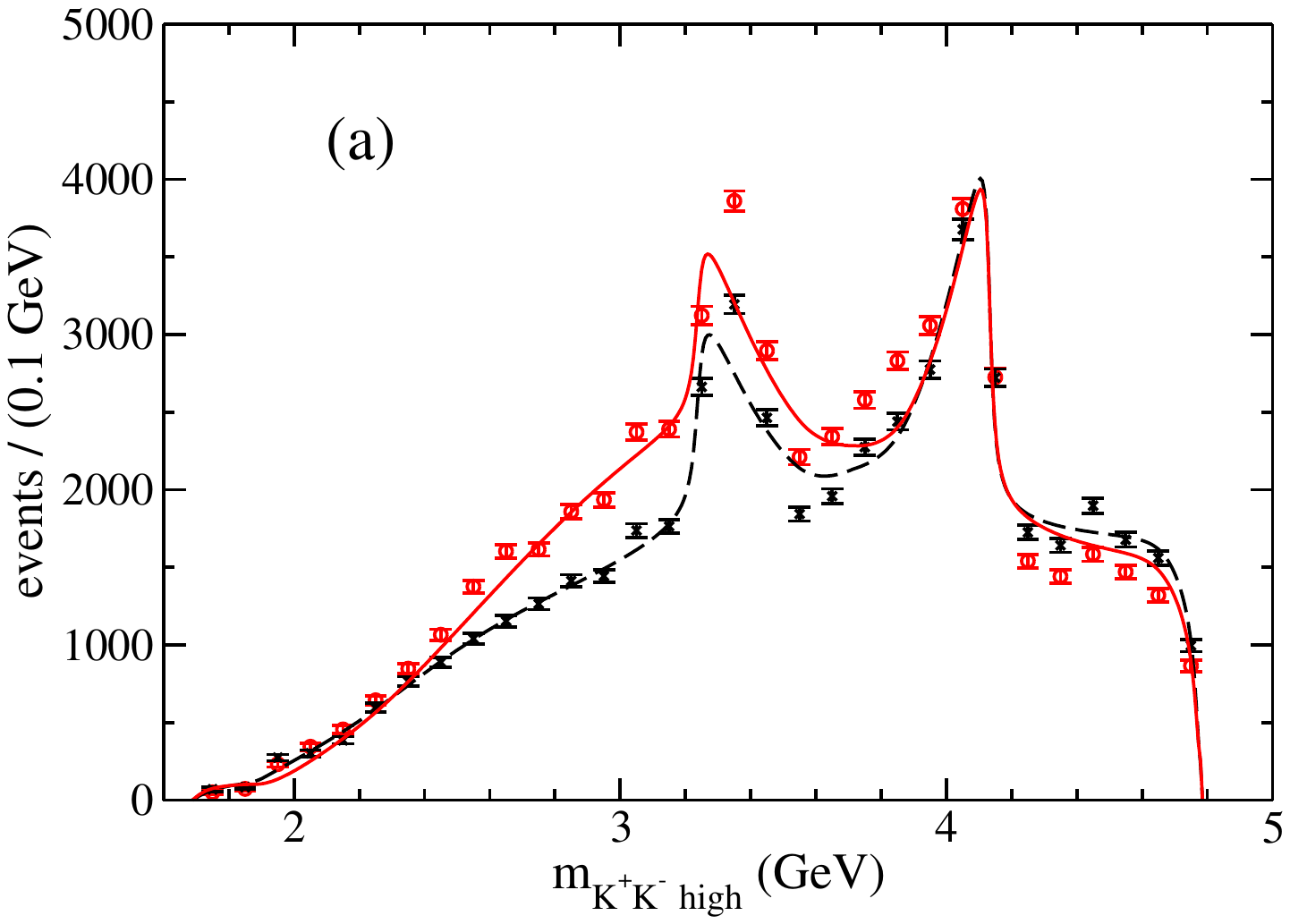}~~
\includegraphics[width=8cm]{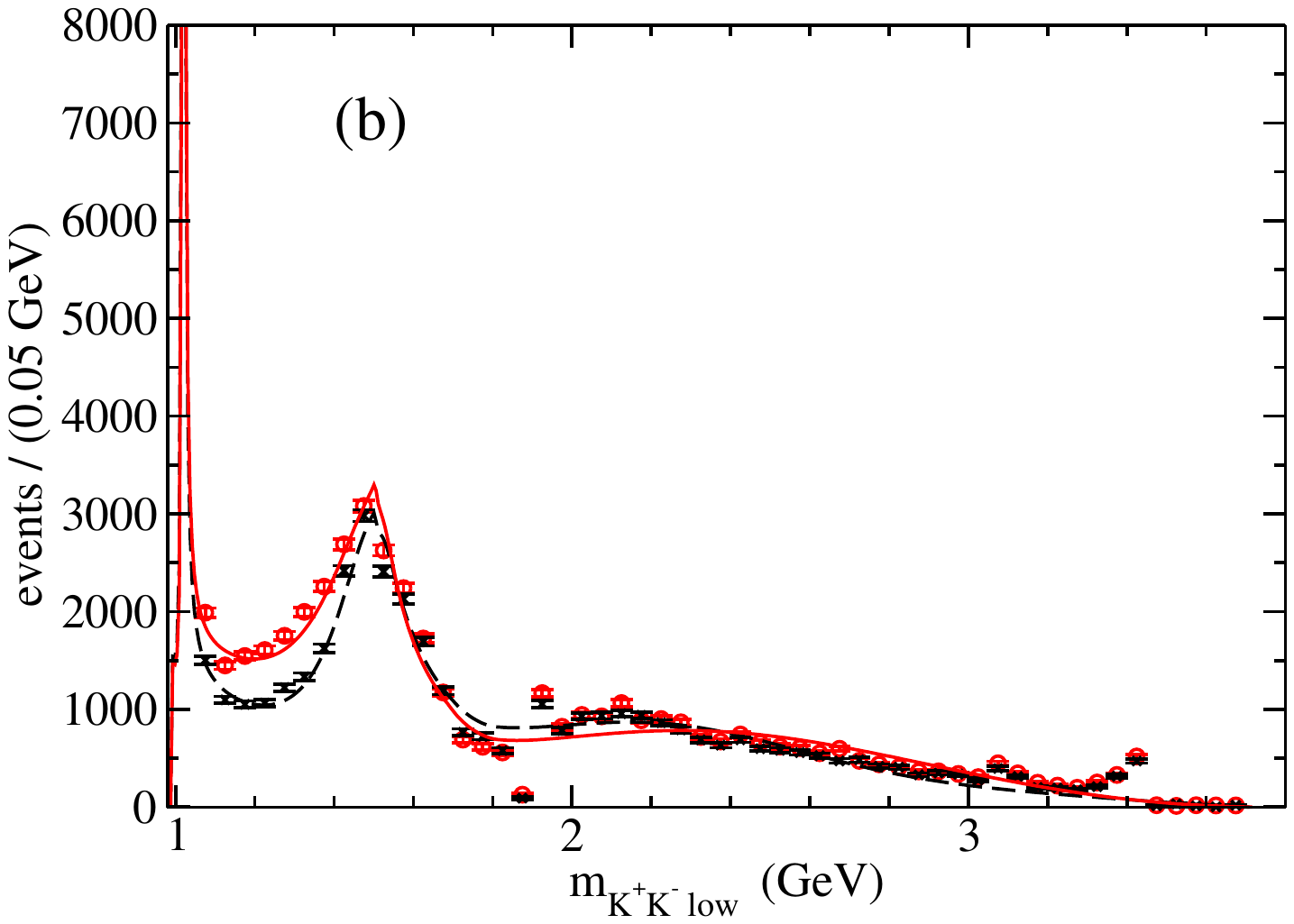}
\end{center}
\caption{\label{fig7}Distributions of the $m_{K^+K^-\,high}$ variable (a) and the $m_{K^+K^-\,low}$ variable (b).
The LHCb Collaboration data for the 
$B^{\pm} \to K^{\pm}K^+K^-$ decays are taken from Fig.~2 of Ref.~\cite{LHCb16}.
The red points correspond to the $B^+$ decays while the black ones to the $B^-$ events.
The continuous red lines are calculated from our model for the $B^+$ decays while the
black dashed ones for the $B^-$ decays.}
\end{figure}

\begin{figure}[h]
\begin{center}
\includegraphics[width=8cm]{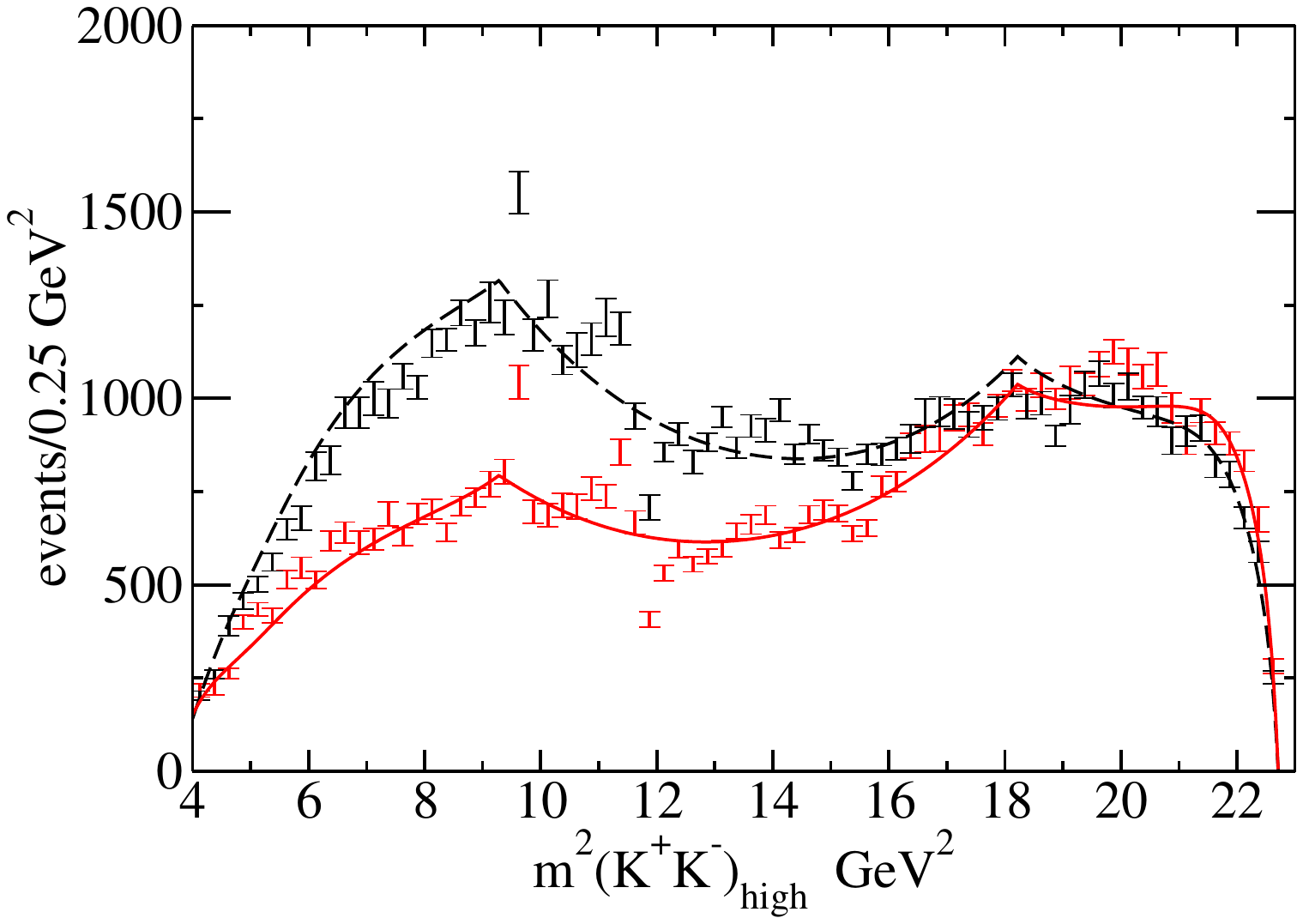}
\end{center}
\caption{\label{fig8}Distributions of the $m_{K^+K^-\,high}$ variable.
The LHCb Collaboration data for the 
$B^{\pm} \to K^{\pm}K^+K^-$ decays are taken from Fig.~7(a) of Ref.~\cite{LHCb23}.
The red points correspond to the $B^+$ decays while the black ones to the $B^-$ events.
The results of our model are represented by the red line for the $B^+$ decays and by the black dashed line for the $B^-$ decays.} 
\end{figure}

The theoretical total branching fractions $Br^{th}(B^+)$ and $Br^{th}(B^-)$
are calculated by the Dalitz plot integration of the differential branching fractions for the $B^{+} \to K^{+} K^+ K^-$ and the 
$B^{-} \to K^{-} K^+ K^-$ decays, respectively.
We have obtained $Br^{th}(B^+)$= 3.48$\times$10$^{-5}$ and $Br^{th}(B^-)$= 3.19$\times$10$^{-5}$.
Then the direct $CP$ asymmetry defined as
\be
\label{ACP}
A_{CP}=\frac{Br^{th}(B^-)-Br^{th}(B^+)}{Br^{th}(B^-)+Br^{th}(B^+)}
\ee
takes the value $A_{CP}$=\,-4.5~\%.
This number can be compared with the experimental value of the $BABAR$ Collaboration
(-1.7$^{+1.9}_{-1.4}\pm$1.4)~\%~\cite{BABAR12} and the LHCb results 
(-3.6$\pm0.4\pm0.2\pm0.7)$~\% from~\cite{LHCb14} and (-3.7$\pm0.2\pm0.2\pm0.3)$~\% from~\cite{LHCb23}.
\begin{table}
\caption{Averages of the branching fractions for the $B^+$ and $B^-$ decays corresponding to separate $S$-, $P$- and $D$\,-wave $K^+K^-$ amplitudes. In the 4th column ratios of these averages to the total averaged branching fraction 
$Br^{th}=3.33\times$10$^{-5}$ are given. The 5th column contains the $CP$ asymmetries.}
\begin{center}
\begin{tabular}{ccccc}
\hline
\hline
Amplitudes & wave& $\frac{1}{2}[Br(B^+)$+$Br(B^-)]$ &ratio to $Br^{th}$ in \% &$A_{CP}$ in \%\vspace*{0.1cm}\\ \hline
$A_1$+$A_2$ &  S &  2.89$\times$10$^{-5}$  &  86.6 & -2.2 \vspace*{0.1cm} \\   
$A_3$+$A_4$ &  P &  1.36$\times$10$^{-5}$&    40.8 & -9.7\vspace*{0.1cm} \\ 
$A_5$+$A_6$ &  D &  0.59$\times$10$^{-6}$ &    1.8 &-0.9\vspace*{0.1cm} \\
\hline
  & sum &  4.31$\times$10$^{-5}$  &   129.2&-4.5 \vspace*{0.1cm} \\
\hline
\hline
\end{tabular} 
\end{center}
\label{branching}
\end{table}

It is interesting to determine contributions to the total branching fraction of the different decay amplitudes grouped according to the relative spin of the $K^+K^-$ pair. 
Thus in Table~\ref{branching} we show the averages of the integrated branching fractions 
for the $B^+$ and $B^-$ decays calculated separately for the $S$- , $P$- and $D$- waves.
Also the ratios to the value of the branching fraction $Br^{th}$ are given.
We see that the $S$-wave amplitudes dominate, the $P$-wave amplitudes are also important, but the $D$-wave ones are small. 
From the numbers shown in this table we can deduce that the interference terms between the $S$- , $P$- and $D$- parts amount to a negative value of -29.2 \%.

We have also studied the interference effects appearing in the contributions to the branching fraction coming from the $S$-wave amplitudes (see Eqs.~(\ref{A1}) and~(\ref{A2})).
Let us denote by $Br_S$ the average of the $S$-wave branching fractions for the $B^+$ and $B^-$ decays.
Then one can separately calculate the contributions to $Br_S$ of the isospin zero
and isospin one $S$-wave amplitudes. 
The isospin zero amplitude is the sum of the $A_1$ amplitude
given by Eq.~(\ref{A1}) and the first term of the amplitude $A_2$ in Eq.~(\ref{A2}) which is proportional to the non-strange kaon form-factor $\Gamma^{n*}_2(s_{23}$).
The isospin one amplitude is proportional to the $G_1(s_{23})$ function seen in Eq.~(\ref{A2}). 
In Table~\ref{BrS} we present the two isospin branching fractions as well as the sizable value of the interference term between these isospin amplitudes.  
In the isospin zero case the amplitude $A_1$ gives $1.44\times 10^{-5}$ contribution while the isospin one term adds $0.58\times 10^{-5}$ to the corresponding branching fraction.
The interference term between these isospin parts of the amplitude is positive and large as it constitutes about 30\% of the total rate in comparison with about 50\% and 20\% relative ratios of the isospin zero and one, respectively.     

\begin{table}
\caption{Contributions of the isospin zero and one $S$-wave amplitudes and its interference to the average $Br_S$ of the $S$-wave branching fractions for the $B^+$ and $B^-$ decays.}
\begin{center}
\begin{tabular}{ccc}
\hline
\hline
     & branching fraction &ratio to $Br_S$ in \% \vspace*{0.1cm}\\ \hline
isospin 0 &  1.44$\times$10$^{-5}$  &  49.8  \vspace*{0.1cm} \\   
isospin 1 &  0.58$\times$10$^{-5}$&   19.9 \vspace*{0.1cm} \\ 
interference term& 0.87$\times$10$^{-5}$ & 30.3 \vspace*{0.1cm} \\
\hline
$Br_S$   &   2.89$\times$10$^{-5}$  &   100.0 \vspace*{0.1cm} \\
\hline
\hline
\end{tabular} 
\end{center}
\label{BrS}
\end{table}

In the last column of Table~\ref{branching} one finds the $CP$ asymmetries. The largest asymmetry is obtained for the $P$-wave amplitudes.
The $CP$ asymmetries for the $S$- and $D$- states integrated over the full Dalitz plot are smaller. 
However, this does not mean that there do not exist regions in the Dalitz plot where the $CP$ asymmetry is large.
Indeed, in 2011 in Ref.~\cite{FKLZ} before a publication of the $BABAR$ ~\cite{BABAR12} and the LHCb data~\cite{LHCb14} we have predicted a large $CP$ asymmetry.
It is particularly well visible in Figs.~\ref{fig5}(a) and ~\ref{fig6}(a) for the $m_{K^+K^-\,low}$ mass range below 1.4 GeV.
This negative asymmetry is larger for the range of the helicity angles 
in which cos$\, \theta_H$ is negative.
This is shown in more detail in Table~\ref{ACP14} where the $B^\pm$ branching fractions and $CP$ asymmetries in the regions of the negative and positive values of
cos$\,\theta_H$ as well as in the full range of helicity angles are presented.
A closer insight in the origin of these large $CP$ asymmetries indicates an essential role of the $\rho(770)$ and the $\rho(1450)$
resonance contributions which are present in the $P$-wave amplitudes $A_3$ and $A_4$ (Eqs.~(\ref{A3}) and (\ref{A4})).
Here the interference effects between the $S$- and $P$-waves are 
important. 

In Fig.~\ref{fig5}(a) one can see a change of sign of the $CP$ asymmetry starting from the $m_{K^+K^-\,low}$ mass exceeding about 1.4 GeV.
This feature of the LHCb data is well reproduced by our model.
The case of cos$\,\theta_H>0$ shown in Fig.~\ref{fig6}(a) for $m_{K^+K^-\,low}> 1.5$ GeV is less clear.
First of all the data for both the $B^+$ and $B^-$ decays have somewhat irregular behaviour at $m_{K^+K^-\,low}$ in vicinity of 1.5 GeV.
One can also notice that the three $B^+$ decay data points at $m_{K^+K^-\,low}$ around 1.6 GeV lie above the correspoding theoretical curve.  
However, an inspection in Fig. 2 of Ref.~\cite{LHCb14} can lead to a guess that this behaviour is related to a kind of interference effects in the Dalitz plot distributions due to a presence of the horizontal band of events related to the 
$\chi_{c0}$ decays into the $K^+K^-$ pairs.
These $\chi_{c0}$ decays cannot influence the distributions of events with cos$\,\theta_H>0$ and we see that in Fig.~\ref{fig5} the mass distributions are more smooth than those seen in Fig.~\ref{fig6}.
In our model we do not describe the decays $B^{\pm} \to K^{\pm} \chi_{c0}$, so some deviation from the data in that rather narrow mass range can be expected but only for cos$\,\theta_H<0$.

\begin{table}
\caption{Branching fractions for the $B^+$ and $B^-$ decays integrated over the 
$m_{K^+K^-\,low}$ mass range between 1.05 GeV and 1.4 GeV and the corresponding $CP$ asymmetries in the three ranges of the helicity angle $\theta_H$.}
\begin{center}
\begin{tabular}{cccc}
\hline
\hline
range of cos$\, \theta_H$ & $Br(B^-)$&$Br(B^+)$ &$A_{CP}$ in \%\vspace*{0.1cm}\\ \hline
cos$\, \theta_H<0$&  2.50$\times$10$^{-6}$  & 4.35$\times$10$^{-6}$ &-26.9 \vspace*{0.1cm} \\   
cos$\, \theta_H>0$&  3.29$\times$10$^{-6}$&  3.65$\times$10$^{-6}$ &-5.2 \vspace*{0.1cm} \\ 
full &  5.79$\times$10$^{-6}$ &  8.0$\times$10$^{-6}$&-16.0\vspace*{0.1cm} \\
\hline
\hline
\end{tabular} 
\end{center}
\label{ACP14}
\end{table}

In Fig.~\ref{fig7} two Dalitz plot projections of the $m(K^+K^-)_{high}$ and 
$m(K^+K^-)_{low}$ variables are given. The theoretical curves describe well the experimental data of the LHCb Collaboration presented in ~\cite{LHCb16}.
Two maxima seen in Fig.~\ref{fig7}(a) correspond mainly to the $B^{\pm} \to \phi(1020) K^{\pm}$ decay events. The same events are grouped in a very narrow peak near 1 GeV in Fig.~\ref{fig7}(b).
The other maximum seen at about 1.5 GeV can be explained as a combined effect of the existence of the three maxima 
present in the strange $\Gamma^{s*}_2(s_{23})$, non-strange $\Gamma^{n*}_2(s_{23})$ scalar kaon form factors and in the transition function $G_1(s_{23})$.
These functions are present in the amplitudes $A_1$ and $A_2$ (Eqs.~(\ref{A1}) and (\ref{A2})).
The three scalar resonances $f_0(1370)$, $f_0(1500)$ and $a_0(1450)$ play a role in an appearence of the above mentioned maximum at 1.5 GeV.
One may also ask as in Ref.~\cite{Zou2021} what is a role of the resonance $\rho(1450)$.
We have performed a special calculation of the $\rho(1450)$ contribution to the height of the $m(K^+K^-)_{low}$ maximum in Fig.~\ref{fig7}(b).
The relative values of this contribution with respect to the experimental numbers of events vary between 3\% and 4\% for the $B^-$ and $B^+$ decays, respectively.
So our conclusion is that the $\rho(1450)$ resonance alone is not responsible for the 
maximum seen at about 1.5 GeV.

In Fig.~\ref{fig8} we show a comparison of the $m^2(K^+K^-)_{high}$ projections with the recent LHCb data (Fig. 7(a) of \cite{LHCb23}). 
These projections correspond to the so-called rescattering region defined as a rather narrow band on the Dalitz plot obtained by limiting the $m^2(K^+K^-)_{low}$ values between 1.1 GeV$^2$ and 2.25 GeV$^2$.
One observes a strong negative $CP$ asymmetry for the $m^2(K^+K^-)_{high}$ values lower than about 17 GeV$^2$. 
The mass projections corresponding to the $B^+$ and $B^-$ decays are rather well described by the model except of a few points at $m^2(K^+K^-)_{high}$ values near 9.6 and 11.7 GeV$^2$ corresponding to the $J/\Psi$ and $\chi_{c0}$ decays.
Let us notice that the negative $CP$ asymmetry effect seen in this figure is also present
in Fig.~\ref{fig6}(a) for the $m(K^+K^-)_{low}$ values smaller than 1.5 GeV.

\section{Summary and final comments}
\label{summary}
We have analysed decays of the $B^{\pm}$ mesons into three charged kaons $K^{\pm}K^+K^-$.
The decay amplitudes have been constructed in the framework of the QCD factorization model and then used to describe the kaon distributions in the whole Dalitz plot.
The model parameters have been simultaneously fitted to the data of the $BABAR$ and LHCb collaborations and a good agreement has been obtained in a description of the 
$K \bar{K}$ effective mass distributions in their full kinematical ranges.
The strong interactions between the $K^+$ and $K^-$ mesons have been taken into account.
These interactions can have a resonant character for lower effective  masses starting from the $K \bar{K}$ threshold.
Many resonances which can decay to $K^+K^-$ pairs in the $S$-, $P$- and $D$-waves give contributions to the decay amplitudes.

At higher effective masses above about 2 GeV the $KK$ interactions may have non-resonant features and in our model we have parameterized them in terms of the polynomials included in the $S$-wave amplitudes.
The Dalitz-plot distributions are dominated by these $S$-wave contributions.
The $P$-wave amplitudes are especially important at low $K^+K^-$ masses with
the $\phi(1020)$, $\rho(770)$ and $\rho(1450)$ resonances interfering strongly with
the $S$-wave amplitudes.
The interference effects depend on the $B$ meson charges leading to the $CP$ asymmetry which varies significantly over the Dalitz-plot. 
Generally the $CP$ asymmetry is negative in the Dalitz-plot range limited from one side by the effective mass $m_{K^+K^-\,low}$ smaller than about 1.6 GeV and by
the $m_{K^+K^-\,high}$ smaller than about 3.7 GeV from the second side.
In other parts of the Dalitz-plot the $CP$ asymmetry is mostly positive and this leads to rather small negative integrated asymmetry equal to -4.5\%.

An important comment should be made here. 
The particle distributions on the Dalitz-plot, the $m_{K^+K^-\,low}$ and $m_{K^+K^-\,high}$ projections and in particular the $CP$ asymmetry distributions cannot be
analysed and understood without taking into account the symmetrization properties of the decay amplitudes as written in Eq.~(\ref{A-sym}). 
This symmetrization leads to a rich structure of the above mentioned functions 
of two variables $s_{23}$ and $s_{12}$.
In particular, one cannot interpret the interference effects of the $S$- and $P$-wave
decay amplitudes using an assumption that the $S$-wave amplitude depends only on the     $s_{23}$ variable and the $P$-wave amplitude is a function of $s_{23}$
multiplied by the cosine of the helicity angle $\theta_H$.
The symmetrization effects, numerically large in the most parts of the Dalitz-plot,
make this assumption not valid.\\   

\newpage
ACKNOWLEDGEMENTS\\

We would like to thank Irina Nasteva for a correspondence 
         concerning the LHCb data.
         We are also grateful to Bachir Moussallam for a calculation of the kaon form factor values and for many useful discussions.

\end{document}